\documentclass[useAMS,usenatbib]{mn2e}
\pdfoutput=1
\usepackage{graphicx,subfigure}
\usepackage[utf8]{inputenc}
\usepackage{amsmath}
\usepackage{amssymb}
\usepackage{units}
\usepackage{booktabs}
\usepackage{array}
\usepackage{multirow}
\usepackage{url}
\usepackage{color}
\usepackage{textcomp}
\usepackage{lscape}
\usepackage[font=footnotesize,labelfont=bf]{caption}
\newcommand{\MSun}{\mbox{${\rm M}_\odot$}}

\def\apgt{\ {\raise-.5ex\hbox{$\buildrel>\over\sim$}}\ }
\def\aplt{\ {\raise-.5ex\hbox{$\buildrel<\over\sim$}}\ }
\def\lt{\ {\raise-.5ex\hbox{$\buildrel>$}}\ }
\def\gt{\ {\raise-.5ex\hbox{$\buildrel<$}}\ }
\def\eqgt{\ {\raise-.5ex\hbox{$\buildrel>\over-$}}\ }
\def\eqlt{\ {\raise-.5ex\hbox{$\buildrel<\over-$}}\ }

\begin{document}
\title{The secular evolution of the Kuiper belt after a close stellar encounter}
  
\author[D. Punzo, R. Capuzzo-Dolcetta, S. Portegies Zwart ]{D. Punzo$^{1}$ $^{2}$, R. Capuzzo-Dolcetta$^{1}$, S. Portegies Zwart$^{3}$\\
$^{1}$ Dep. of Physics, Sapienza, University of Roma, P.le A. Moro
  1, Roma, Italy\\
$^{2}$ Kapteyn Institute, Rijksuniversiteit, Landleven 12, 9747AD
  Groningen, Netherlands\\
$^{3}$ Leiden Observatory, Leiden University, P.O. Box 9513, 2300 RA Leiden, The Netherlands }

\maketitle

MNRAS Accepted 2014 August 11th. Received 2014 July 14th; in original form 2014 March 21th.

\begin{abstract}

We show the effects of the perturbation caused by a passing by star on
the Kuiper belt objects (KBOs) of our Solar System.
The dynamics of the Kuiper belt (KB) is followed by direct $N$-body simulations.
The sampling of the KB has been done with $N$ up to $131,062$,
setting the KBOs on initially nearly circular orbits distributed in a
ring of surface density $\Sigma \sim r^{-2}$.    
This modelization allowed us to investigate the secular evolution
of the KB upon the encounter with the perturbing star.
Actually, the encounter itself usually leads toward eccentricity
and inclination distributions similar to observed ones, but tends also to
excite the low-eccentricity population ($e\aplt 0.1$ around
$a\sim 40$\,$\mathrm{AU}$ from the Sun), depleting this region of low
eccentricities. The following long-term evolution shows a ``cooling" of
the eccentricities repopulating the low-eccentricity area.
In dependence on the assumed KBO mass spectrum and sampled number of bodies,
this repopulation takes place in a time that goes from 0.5\,Myr to 100\,Myr.
Due to the unavoidable limitation in the number of objects in our long-term simulations
($N \leq 16384$), we could not consider a detailed KBO  mass spectrum, accounting for low mass 
objects,  thus our present simulations are not reliable in constraining correlations
among inclination distribution of the KBOs and other properties, such as their
size distribution. However, our high precision long term simulations are a starting point
for future larger studies on massively parallel computational platforms which will provide a 
deeper investigation of the secular evolution ($\sim 100\,$Myr) of the KB over its whole mass 
spectrum.
\end{abstract}


\section{Introduction}

The Solar System is hedged by a ring composed of a huge number of
small bodies: the Edgeworth-Kuiper belt \citep{Jewitt} (hereafter
briefly called Kuiper belt, or KB).  The Kuiper belt bears the
signature it the early evolution of the Solar System, and a contains
records of the end-state of the accretion processes occurred in that
region. Therefore, the knowledge of the history of the Kuiper belt
objects (KBOs) is relevant to be able to develop a full consensus of
the formation of the Solar System.

The majority of the KBOs are located between about 30\,$\mathrm{AU}$ 
and 90 $\mathrm{AU}$ from the Sun, but most are around the 2:3 
resonance with Jupiter, at 39.5\,$\mathrm{AU}$ and at its 1:2 
resonance, roughly around 48\,$\mathrm{AU}$. The total
mass is estimated from $0.01$ to $0.1$ M$_\oplus$ \citep{Luu}. There
are several, indirect, arguments suggesting that this is just a small
fraction of its initial mass because most of it has been lost (see 
\cite{Kenyon}). The size distribution of the KBOs is,
usually, assumed as a power law $dn/dR = A R^{-q}$, where $A$ and $q$
are constants. The $q$ exponent is estimated $\sim 4.0 \pm 0.5$
\citep{Ber,Fraser}. For a more detailed description of the
Kuiper belt we refer, e.g., to \cite{Luu}.

The KB has a bimodal inclination distribution
resulting of two separate populations \citep{Brown}. The
\emph{dynamically cold} population refers to objects 
moving on almost planar orbits with relatively low inclinations (up to about
10$^\circ$) respect to the ecliptic. On the other side, the
dynamically hot population is characterized by highly inclined orbits (up
to 40$^\circ$) with respect to the ecliptic. Note that these two populations are 
different from what we call, in this paper, the \emph{low-eccentricity} population, 
which are objects on nearly circular orbits (orbital eccentricities $< 0.1$) and the 
\emph{high-eccentricity} population (eccentricities $\gtrsim$ 0.1. 

\begin{figure}
\centering
\subfigure{
\includegraphics[scale=0.3]{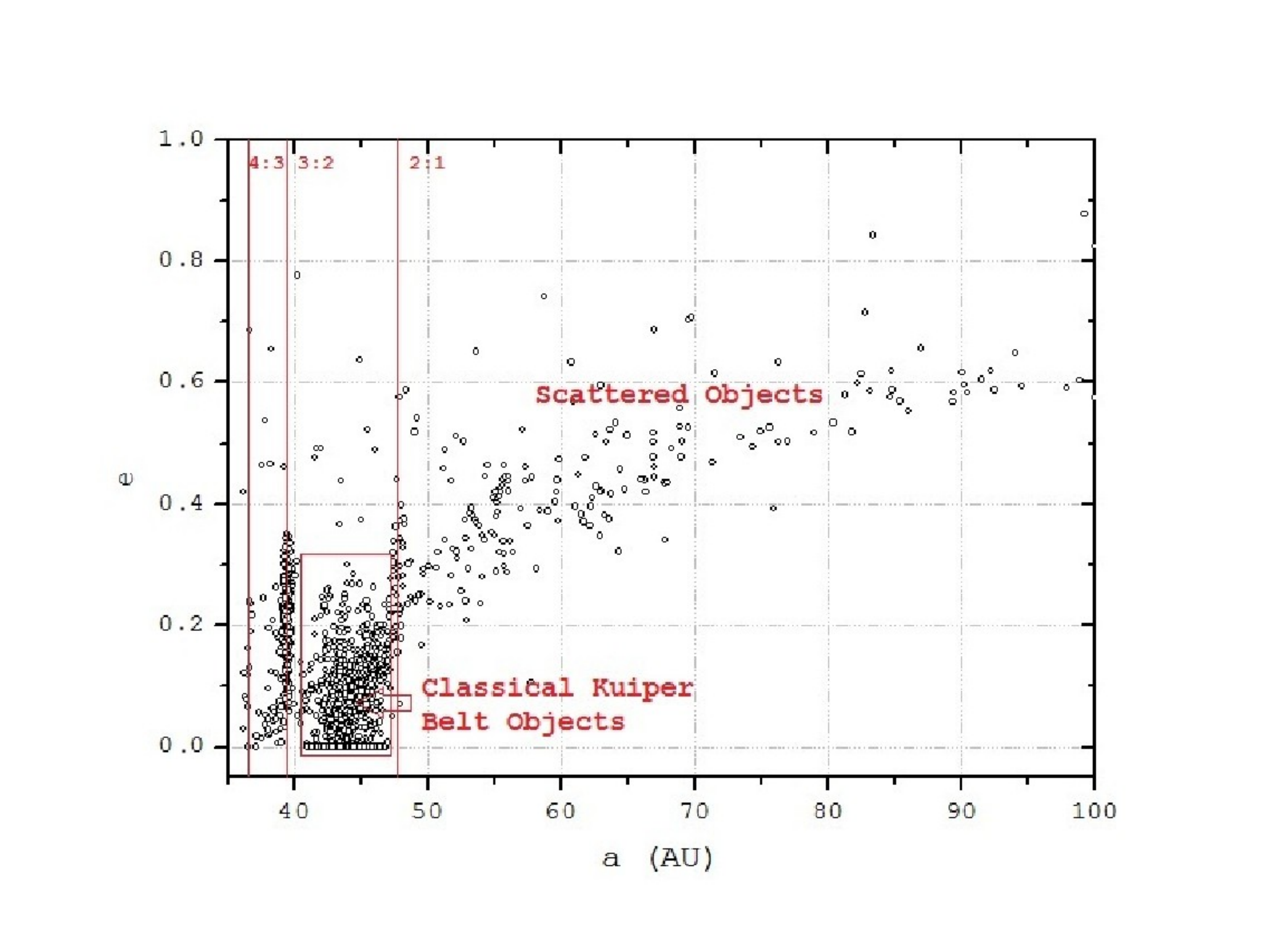}
}
\subfigure{
\includegraphics[scale=0.3]{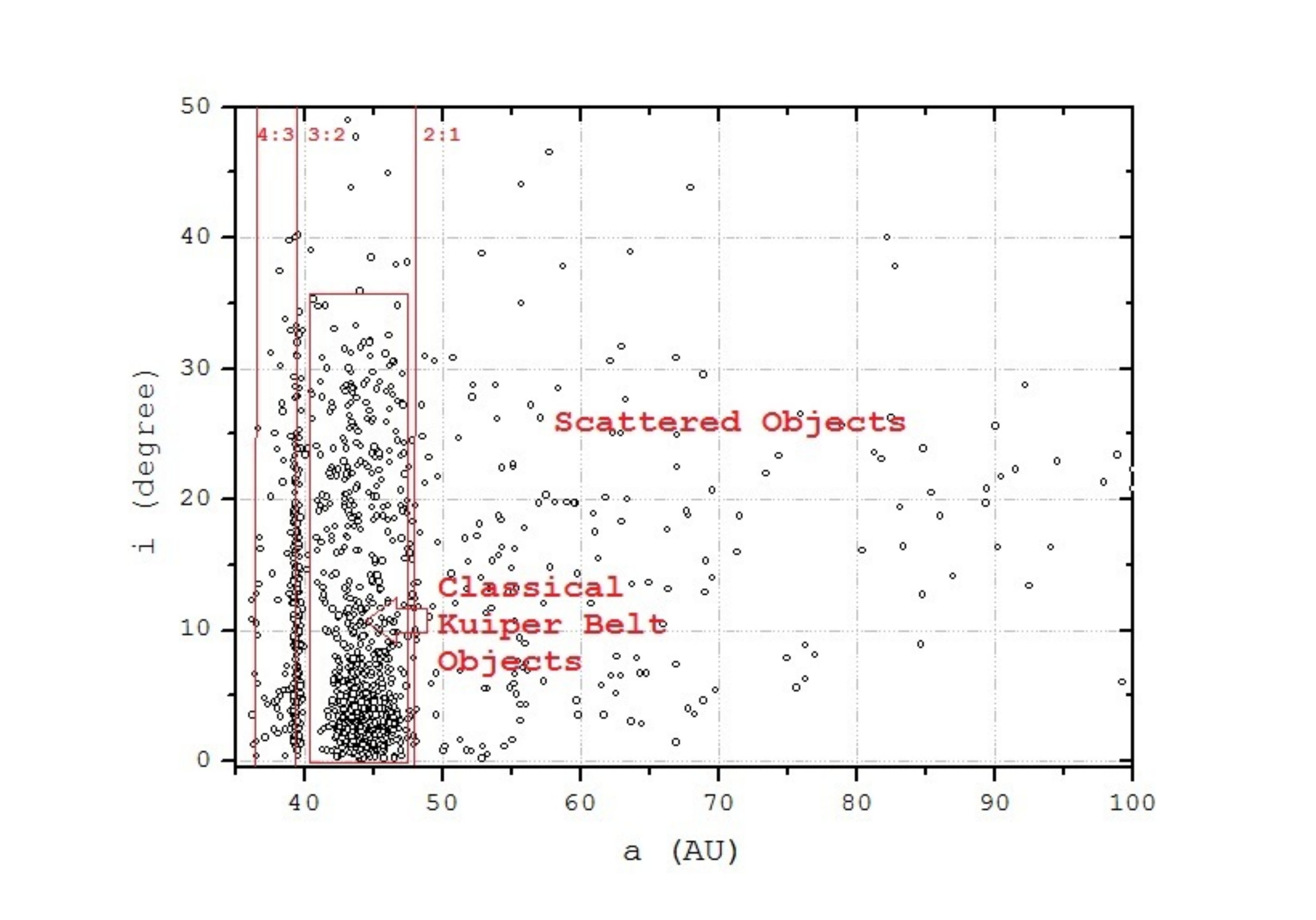}
}
\caption{distribution of the eccentricities (top panel) and 
inclinations (bottom panel) as a function of semi-major axis of the
observed KBOs; data are from the Minor Planet Center \citep{mpc}. The
vertical lines highlight the main resonances.}

\label{fig:oss}
\end{figure}

In Fig. \ref{fig:oss} the eccentricities and inclinations are plotted
as function of the semi-major axis for KBOs observed from the Minor
Planet Center (MPC) \citep{mpc} which is the center of the Smithsonian
Astrophysical Observatory (SAO) dedicated to tracking, monitoring,
calculating and disseminating data from asteroids and comets.

The KBOs have been sub-categorized in three groups:
\begin{enumerate}
\item classical KBOs ($42$ $< a < 49$ , $\langle e \rangle
  \simeq 0.09$, $\langle i \rangle \simeq 7^{\circ}$);
\item scattered KBOs ($a > 30$ , $\langle e \rangle \simeq 0.49$,
  $\langle i \rangle \simeq 14^{\circ}$ );
\item  main resonant KBOs :
\begin{itemize}
\item 4:3 resonance ($a \simeq36,4$, $\langle e \rangle \simeq
  0.22$, $\langle i \rangle \simeq 8^{\circ}$);
\item 3:2 resonance, Plutino's ($a \simeq39.4$, $\langle e
  \rangle \simeq 0.36$,$\langle i \rangle \simeq 13^{\circ}$);
\item 2:1 resonance ($a \simeq47.8$, $\langle e \rangle \simeq
  0.14$, $\langle i \rangle \simeq 10^{\circ}$).,
\end{itemize}
\end{enumerate}   

where semimajor axes, $a$, are in $\mathrm{AU}$.

These sub-populations have been explain through a phase of planet
migration and a phase of clearing of the environment during the
evolution of the early Solar System \cite{Mal1, Mal2}. In the latter
phase the resonance population was formed by sweeping resonance
capture in which the Jovian planets withstand considerable orbital
migration as a result of encounters with residual planetesimals. While
Neptune moved outwards, a small body like Pluto in an initially
circular orbit could have been captured into the 3:2 resonance.  The
high orbital eccentricity would subsequently be induced by repeated
orbital crossings with Neptune.

Many others studies have attempted to better understand the properties
of the KBOs. \cite{Gomes} investigated how the outward
migration of Neptune, as proposed by \cite{Mal1, Mal2}, could have
scattered objects from $25$ $\mathrm{AU}$ onto high-$i$ orbits leading to the
current classical Kuiper belt region. He concluded that the high-$i$
population was formed closer to the Sun and brought into the classical
Kuiper belt during planetary migration, whereas the cold population
represents a primordial, relatively undisturbed population. This also
led to the speculation that other mechanisms, such as planetary
migration, have been the cause of the correlation between inclinations
and colors in the classical Kuiper belt rather than environmental
effects like the collisions among the KBOs (see \cite{Dore}). 
Detailed discussions about the correlation of the inclination with the 
color, size and binary of the KBOs are given by \cite{Levi1, Bru, 
Noll,Volk}. More recently a model, called the Nice model, has been 
proposed \citep{Levi}, which argues that the giant planets migrated 
from an initial compact configuration into their present orbits, long 
after the dissipation of the initial protoplanetary gas disk. The Nice 
model seems to provide a acceptable explanation for the formation of 
the classical and scattered populations, and for the correlation 
between inclinations and colors (for more detail see \cite{Levi}).  The 
Nice model, however, predicts a higher eccentricities in classical KBO 
orbits than is observed.

An interaction between a passing field star and the the Solar System
could also be responsible for some of the orbital families observed in
the KBO, which is the main topic of this paper.

\section{The fly-by star perturbation and the $N$-body scheme}

An encounter between a passing star and the Solar System is quite
likely, considering that the Solar System was probably formed in an
open star cluster \citep{simon1}.  The hypothesis of a closely passing
star has been hypothesized before, and used to explain KBO families
\citep{Ida1,Ida2,Ida3, Melita, Malberg}.  The cost of such calculations,
however, prevented earlier research on the secular
evolution of the KBO by mean of high resolution simulations.

We focus our attention on the investigation of the effects of the long-term
evolution of the Kuiper belt after a close stellar encounter on the structure of the 
Kuiper belt.
We adopt a direct $N$-body treatment in which the mutual, pair-wise,
interactions between KBOs, planets and stars are taken into account
self-consistently. Due to the computational expense of this method we
are limited to about $131,072$ total bodies.

We modeled the early Solar System as composed by the Sun, the eight
major planets, Pluto and the Kuiper belt.  Each object was considered
a point-mass; we did not account for collisions.  The KBOs were
initially moving in circular orbits in a flat ring in the plane of the
ecliptic. This corresponds to an initially cold population, without a
$z-$component in their motion. We adopted a surface density $\Sigma
\propto r^{-2}$, where $r$ is the heliocentric distance (see
\cite{Holman}).

We studied two possible configurations:
\begin{itemize}
\item (i) model A, with a radial extension in the range from $42$ $\mathrm{AU}$ 
to   $48$ $\mathrm{AU}$ and four different values of the total mass of the KB, 
$M   = 3, 6, 1, 30$ M$_\oplus$;
\item (ii) model B, with a radial extension in the range from $42$ $\mathrm{AU}$ 
to   $90$ $\mathrm{AU}$ and a total mass $M = 30$ M$_\oplus$.
\end{itemize}

\begin{figure}
\subfigure{
\includegraphics[width=0.50\textwidth]{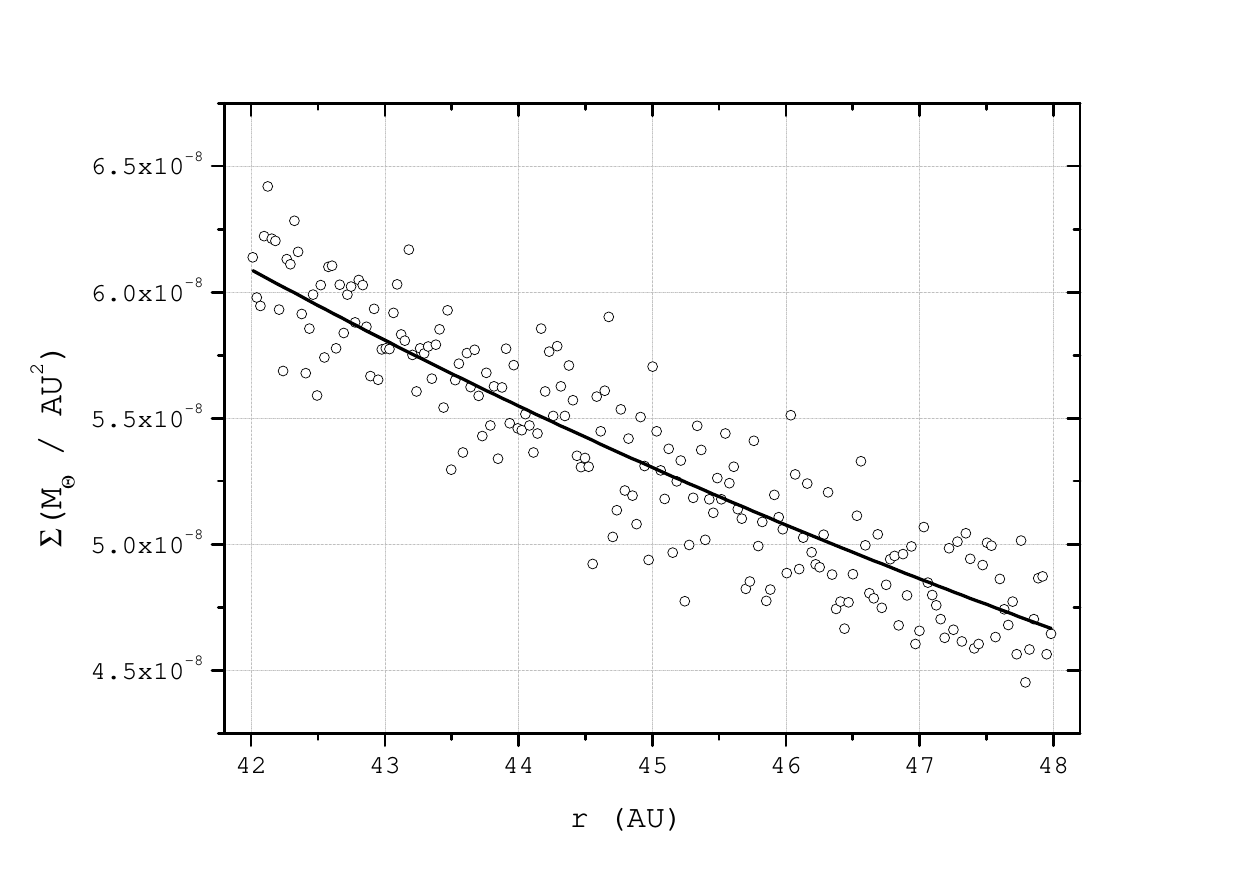}}
\hspace{0mm}
\subfigure{
\includegraphics[width=0.50\textwidth]{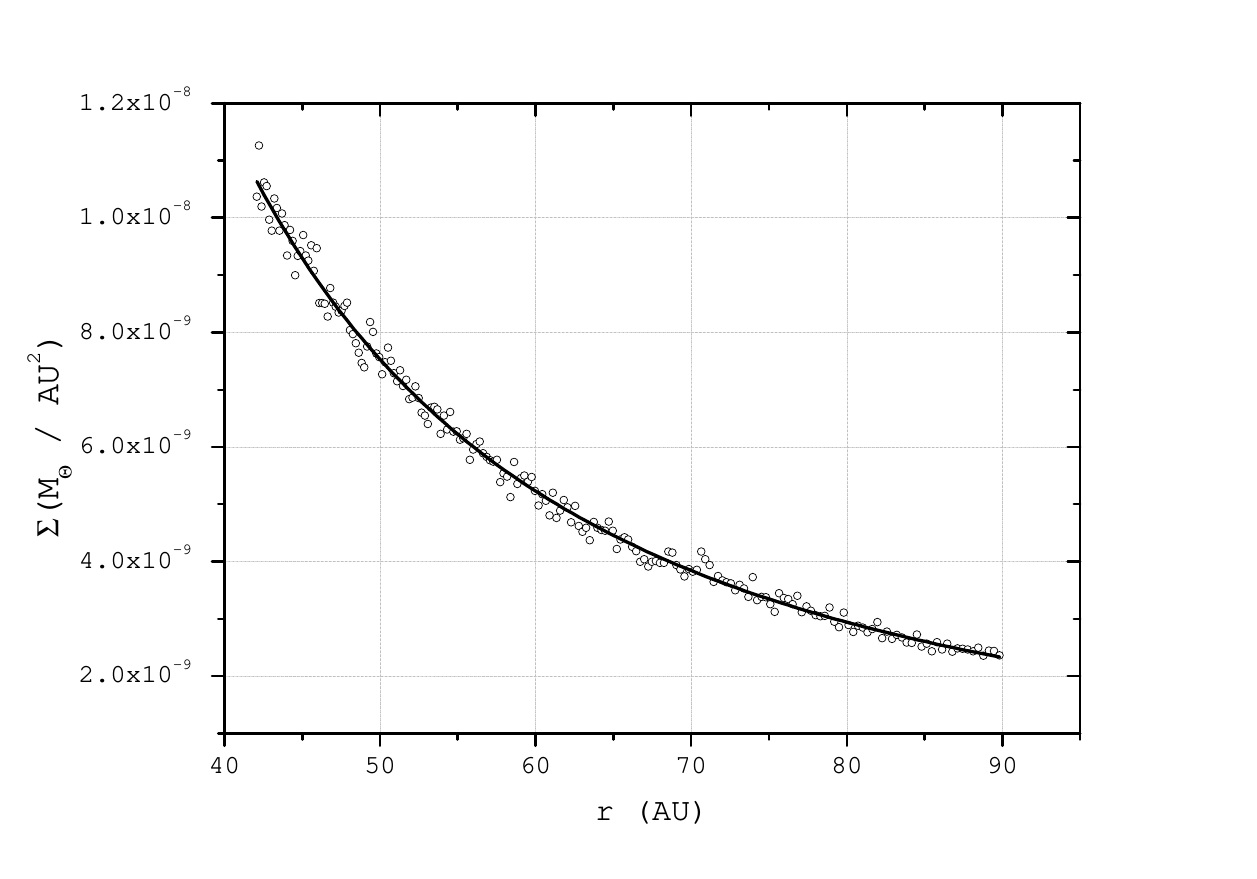}}
\caption{The initial surface density $\Sigma$ of model A, with $M = 30
  $M$_\oplus$ (top panel) and for model B (bottom panel).}
\label{fig:dens}
\end{figure}

The mass function of the KBOs was derived from the conversion in mass
of the size ($R$) distribution, assuming a constant KBO density
(i.e. $\rho \sim 10^3$ kg/m$^{3}$), which results in 
$dm/dR \propto \rho R^{-2.0 \pm 0.5}$ with a cut at
$m_{min} = \frac{4}{3} \pi \rho R^3  \approx 7.0 \; 10^{-13} $M$_\oplus$ 
(corresponding to $R=1$ km) and $m_{max}\approx 7.0 \;
10^{-4} $M$_\oplus$ (corresponding to $R=10^3$ km).  In
Fig. \ref{fig:dens} we present the surface density of the models A and
B as a function of the heliocentric distance for KBOs with the same
individual mass, $m = M / N$. Here $M$ is the total mass of the Kuiper
belt and $N$ the number of KBOs. With this choice we have a good
sampling of the KB without an exceedingly large number ($N> 10^6$) of
particles.

We integrate the equations of motion by direct summation $N$-body 
codes running on Graphics Processing Units (GPUs). For the 
gravitational $N$-body problem these accelerators give a manifold speed
increase with respect to code running on CPU 
\citep{nyland,Simon,Jeroen}. Parallel computers equipped
more than one hundred GPUs have been utilized for various studies 
\citep{HiGPUs, GPUcomp,Berczik1,Berczik2} have been run efficiently in parallel to 
provide the computational power necessary to perform direct many body
simulations. Access to such large GPU-equipped supercomputers,
however, is not easy, in particular when the computations required a
considerable fraction of the available hardware. We therefore mainly
ran our simulation on the Little Green Machine, a at the Sterrewacht
Leiden built dedicated GPU-equipped supercomputer, specifically built
for performing GPU-related calculations.  Even with this machine, we
had to limit the number of bodies to about a hundred thousand, but we
performed several simulations (of models A and B) for each realization
of the initial conditions in order to assure that the results of
our calculations were not a statistical anomaly.  Recently
\cite{Tjarda} demonstrated that performing multiple simulations with
the same initial conditions provide a statistically correct sampling
of the real solutions.  In these runs we varied the mass, impact
parameter and the inclination of the incoming star.

\begin{equation}  \label{eq:init}
\begin{cases}
\begin{array}{cc}
M_{\star} = [0.5;1;2] {\text M}_{\odot}\\
x=500 & v_{x,\infty}\simeq -3, \\
y=b\cos\theta & v_{y,\infty}=0, \\
z=b\sin\theta & v_{z,\infty}=0,
\end{array} 
\end{cases}
\end{equation}

where $x,y,z$ are in $\mathrm{AU}$ and velocities in km/s.
The system of reference was centered on the Sun, and the impact 
parameters $b$ and inclination $\theta$ characterize the orbit of 
the encountering star. The incoming star was placed in a ring of 
radius $b$ at a distance of 500\,$\mathrm{AU}$ in the $x$-direction parallel 
to the $yz$ plane. In Tab.  \ref{tab:tabcond} we  present the 
initial conditions for our simulations.
In order to have a full coverage of the parameter space,
$v_{y,\infty}$ and $v_{z,\infty}$ should be varied as free parameters. 
However, a systematic set of $N$ body simulations is computational expensive 
(at least when considering $N$ large enough to guarantee a good sampling) forced us 
to reduce the investigation in the parameter space. Consequently, we considered that 
the most relevant thing to do was exploring the role of the  initial $yz$ spatial 
coordinates.
Actually, the variation of two free parameters are enough for exploring encounters 
with different strength (see Sect. \ref{resu}). 
Of course, a more extended study of the other free parameters could allow a wider 
comprehension of the role of stellar encounters on the KB structure.

\begin{table*}
\centering
\begin{minipage}{140mm}
\begin{tabular}{c|cccc|ccccc}
\hline 
\hline 
 index & $b$ & $\theta$ & $y(\mathrm{AU})$ & $z(\mathrm{AU})$ &  index & $b$ & $\theta$ & $y(\mathrm{AU})$ & 
 $z(\mathrm{AU})$ \\ 
\hline 
 1 & 140 & 90 & 0.000 & 140.000 &  33 & 200 & 30 & 173.205 & 100.000 \\
 2 & 140 & 100 & -24.311 & 137.873 & 34 & 200 & 60 & 100.000 & 173.205 \\
 3 & 140 & 110 & -47.883 & 131.557 & 35 & 200 & 70 & 68.404 & 187.939 \\
 4 & 150 & 30 & 129.904 & 75.000 &  36 & 200 & 75 & 51.764 & 193.185 \\
 5 & 150 & 60 & 75.000 & 129.904 &  37 & 200 & 80 & 34.730 & 196.962 \\
\hline 
 6 & 150 & 90 & 0.000 & 150.000 &  38 & 200 & 90 & 0.000 & 200.000 \\
 7 & 150 & 100 & -26.047 & 147.721 &  39 & 200 & 105 & -51.764 & 193.185 \\
 8 & 150 & 110 & -51.303 & 140.954 &  40 & 200 & 120 & -100.000 & 173.205 \\
 9 & 150 & 120 & -75.000 & 129.904 &  41 & 200 & 135 & -141.421 & 141.421 \\
 10 & 150 & 150 & -129.904 & 75.000 &  42 & 200 & 150 & -173.205 & 100.000 \\
\hline 
 11 & 160 & 90 & 0.000 & 160.000 &  43 & 212.5 & 60 & 106.250 & 184.030 \\
 12 & 160 & 100 & -27.784 & 157.569 &  44 & 212.5 & 75 & 54.999 & 205.259 \\
 13 & 160 & 110 & -54.723 & 150.351 &  45 & 212.5 & 90 & 0.000 & 212.500 \\
 14 & 170 & 90 & 0.000 & 170.000 &  46 & 212.5 & 105 & -54.999 & 205.259 \\
 15 & 170 & 100 & -29.520 & 167.417 &  47 & 212.5 & 120 & -106.250 & 184.030 \\
\hline 
 16 & 170 & 110 & -58.143 & 159.748 &  48 & 212.5 & 135 & -150.260 & 150.260 \\
 17 & 170 & 90 & 0.000 & 170.000 &  49 & 212.5 & 150 & -184.030 & 106.250 \\
 18 & 170 & 100 & -29.520 & 167.417 &  50 & 225 & 30 & 194.856 & 112.500 \\
 19 & 170 & 110 & -58.143 & 159.748 &  51 & 225 & 60 & 112.500 & 194.856 \\
 20 & 175 & 30 & 151.554 & 87.500 &  52 & 225 & 75 & 58.234 & 217.333 \\
\hline  
 21 & 175 & 60 & 87.500 & 151.554 &   53 & 225 & 90 & 0.000 & 225.000 \\
 22 & 175 & 70 & 59.854 & 164.446 &  54 & 225 & 105 & -58.234 & 217.333 \\
 23 & 175 & 80 & 30.388 & 172.341 &  55 & 225 & 120 & -112.500 & 194.856 \\
 24 & 175 & 90 & 0.000 & 175.000 &  56 & 225 & 135 & -159.099 & 159.099 \\
 25 & 175 & 120 & -87.500 & 151.554 &  57 & 225 & 150 & -194.856 & 112.500 \\
\hline 
 26 & 175 & 150 & -151.554 & 87.500 &  58 & 237.5 & 60 & 118.750 & 205.681 \\
 27 & 180 & 70 & 61.564 & 169.145 &  59 & 237.5 & 75 & 61.470 & 229.407 \\
 28 & 180 & 80 & 31.257 & 177.265 &  60 & 237.5 & 90 & 0.000 & 237.500 \\
 29 & 180 & 90 & 0.000 & 180.000 &  61 & 237.5 & 105 & -61.470 & 229.407 \\
 30 & 190 & 70 & 64.984 & 178.542 &  62 & 237.5 & 120 & -118.750 & 205.681 \\
\hline 
 31 & 190 & 80 & 32.993 & 187.113 &  63 & 237.5 & 135 & -167.938 & 167.938 \\
 32 & 190 & 90 & 0.000 & 190.000 &  64 & 237.5 & 150 & -205.681 & 118.750 \\
\hline
\hline 
\end{tabular}
\caption{The initial impact parameter $b$ and inclination $\theta$,
  and position in $y$ and $z$ of the incoming star for each of the 64
  simulations performed.  The system of reference is centered at the
  Sun.}
\label{tab:tabcond}
\end{minipage}
\end{table*}

Calculations were performed using the direct summation code {\tt
  HiGPUs} \citep{HiGPUs}, which is publicly available via the
Astronomical Multipurpose Software Environment (AMUSE) 
\citep{amuse1, amuse2, simon2}.  

This code uses its own kernels to implement at best a 6th-order 
Hermite's integrator \citep{nitadori} with block time-steps 
 \citep{ars} method.

We tested the accuracy of {\tt 
HiGPUs} in getting the results of interest here through comparison with two 
symplectic $N$-body codes, {\tt NBSymple} \citep{NBSymple}, which is based on a 
symplectic second and sixth order method for the time integration of 
the equations of motion and {\tt HUAYNO} 
\citep{HUYANO}, which uses recursive-Hamiltonian splitting to generate 
multiple-timestep integrators that conserve momentum to machine 
precision. The comparison indicates as fully reliable 
the simulations done with the (much faster) {\tt HiGPUs} code.

All the simulations were performed using a softening parameter,
$\epsilon$, in the pairwise Newtonian potential $U_{ij} \propto
\sqrt{r_{ij}^2+\epsilon^2}$, where $r_{ij}$ is the $i-th$ to $j-th$ 
particle distance. The $\epsilon$ value was set to 
 $4\times 10^{-4}$ $\mathrm{AU}$, which is $\sim 1500$ times 
smaller than the initial average distance to the nearest neighbour in our sampling, 
and $\sim 60$ times bigger than the radius of Pluto. This choice guarantees the 
preservation of the newtonian behaviour of the interobject force while keeping under 
control spurious fluctuations over the mean field (see following Subsect. 
\ref{softening}).
The maximum time step for the hierarchical block time steps was $\sim 0.02$ $yr$.  
The energy conservation was checked along the system evolution by its 
fractional time variation defined as

\begin{equation}
\left|\frac{\Delta E}{E}\right|  = \left| \frac{E(t)-E(0)}{E(0)}\right|.
\end{equation}
At the end of the simulations it was always below the value $10^{-7}$,
which is more than sufficient to assure that we statistically
correctly sample the result of a true (converged) solution to the
$N$-body problem \citep{Tjarda}.

\subsection{The role of softening}\label{softening}

The real KB is likely composed of various thousands objects. The study of the secular evolution of the KB with a 
high-precision, direct summation, $N$-body code after the encounter with a passing-by star is out of reach with our available hardware.
For this reason, to represent the KBOs we limited to values of $N$ just below $10^4$, taking as reference value 
$N=2^{13}-10 = 8182$ (ten bodies represent the incoming star, the Sun and the planets) which
showed to be a good compromise between accuracy (resolution) and computational speed.
Of course, to give physical reliability to our results obtained by such subsampling, we needed the introduction 
of a softening parameter ($\epsilon$, as described above) whose size must be calibrated. Actually, the role of 
softening parameter in the N-body simulation is a long, highly debated question. It is well known that a softening 
parameter in the Newtonian particle-particle force has the double role of i) avoiding the ultraviolet singularity 
in the closest interaction and ii) reducing the spurious granularity effects induced by the use of a sub sampled 
N-body set of particles to reproduce the evolution of a large stellar system.
The choice of the softening length, $\epsilon$, is characterized by a proper balance between the width of the 
softening length, to be small enough to preserve the Newtonian behaviour and, at the same time, large enough 
to avoid spurious collisionally in the evolutionary behaviour of the system.
To fulfil the second requirement above, $\epsilon$ is necessarily much larger than the average KBO radius but 
this is not a serious issue because the average close neighbour distance, $\langle d_{cn} \rangle$, in the 
simulated KB system is $\simeq 0.6$ $\mathrm{AU}$ $>>$ the average KB radius. On the other hand, the first requirement above 
(preserve Newtonian behaviour of the force) requires an $\epsilon$ sufficiently smaller than  $\langle d_{cn} \rangle$. Actually, 
we tested the simulations with two values of the softening ($4\times 10^{-4}$ $\mathrm{AU}$ and $4\times 10^{-5}$ $\mathrm{AU}$ 
for $\epsilon$), which are both significantly smaller than $\langle d_{cn}  \rangle$ which is  $\simeq 0.6$ $\mathrm{AU}$ for $N=8182$. 
As additional, practical, confirmation that the range of $\epsilon$ explored corresponds to reliable results, we saw that overall results 
remain almost unchanged with the two different choices for the $\epsilon$ value. 
So, we feel quite confident that our $N$ body results are solid in stating about KB secular evolution after the stellar encounter.

\section{Results}\label{resu}

Here we report on the results of the simulations using our two initial
conditions, model A and model B.

\subsection{Model A}

In model A we adopted a radial extension of the KBO between 42\,$\mathrm{AU}$ to
48\,$\mathrm{AU}$ using a total mass of $M = 1$, 3, 6 and 30\,M$_\oplus$.  The
mass and encounter parameters of the incoming star are presented in
Tab.\,\ref{tab:tabcond}. The simulations are carried out until the
perturbation induced by the passing star is negligible, even compared
to the inter KBO forces (the gravitational contribute on the total 
force on a generic KBO due to the passing star is 
five magnitude lower respect the Sun and one magnitude comparing with 
the nearest KBO neighbour).

\begin{figure*}
\centering
\subfigure{
\includegraphics[width=0.4\textwidth]{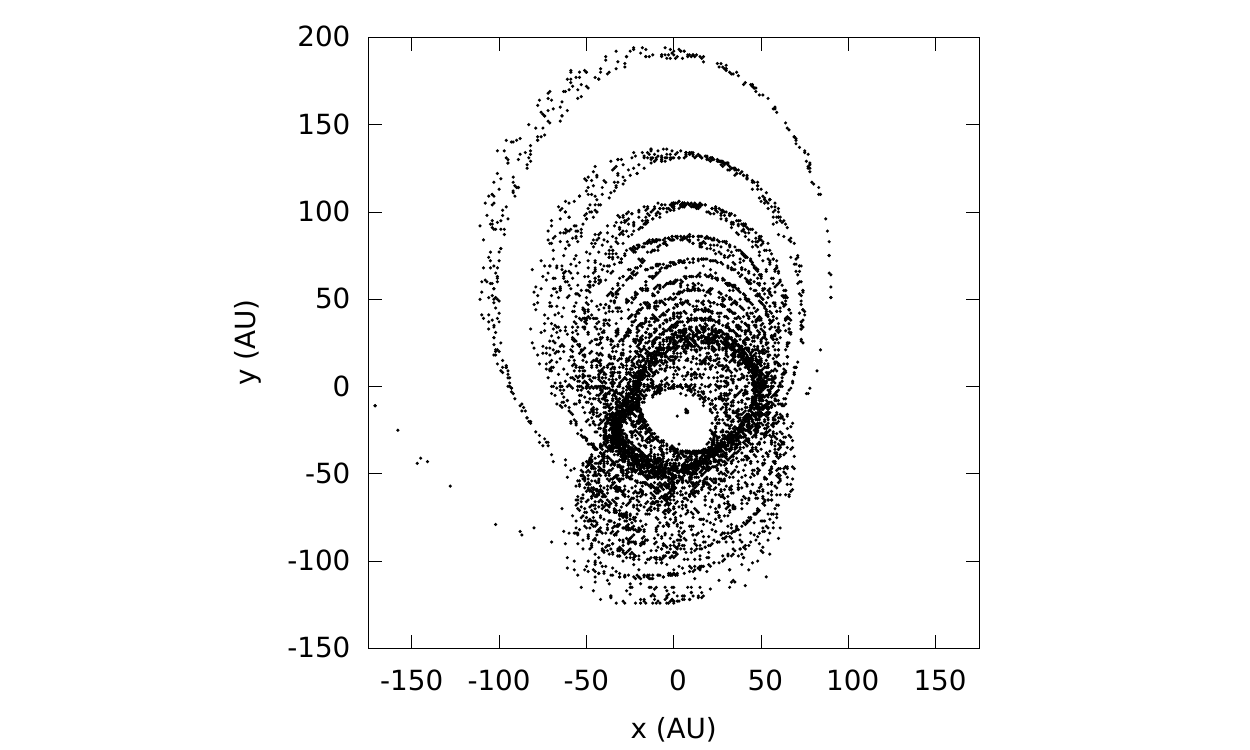}}
\subfigure{
\includegraphics[width=0.4\textwidth]{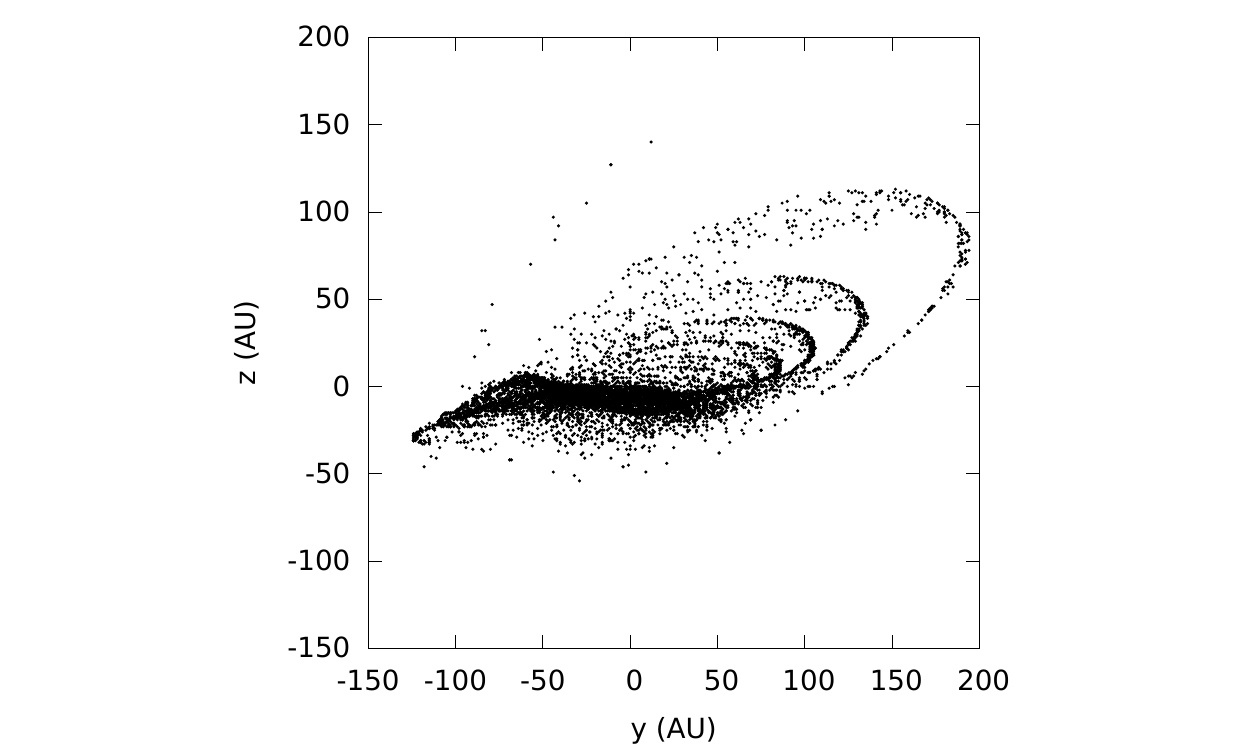}}
\caption{distribution of the $N=8182$ Kuiper belt objects in the $xy$
  plane (left panel) and in the $yz$ plane (right panel) after an
  encounter with a 1\,\MSun\, star (at $T = 3.2 \times 10^3$\,yr) with
  parameters $b = 175$ $\mathrm{AU}$ and $\theta = 90^{\circ}$.}
\label{fig:plane}
\end{figure*}

In Fig. \ref{fig:plane} we give the distribution of the KB as obtained
by model A. The figure presents the projection of the sampled system
onto the $xy$ and the $yz$ planes $\sim 2500$\,yr after the closest 
approach between the Sun and the perturbing star.  Some KBOs were 
scattered out to a heliocentric distances exceeding 200\,$\mathrm{AU}$, and with 
very high eccentricities, but the majority ($\sim 73\%$) of objects 
remains bound. Moreover the KBOs are distributed in over densities 
triggered by the passing of the star which are resonances 
due to the planets contribution. 

\begin{figure*}
\centering
\includegraphics[width=1.0\textwidth]{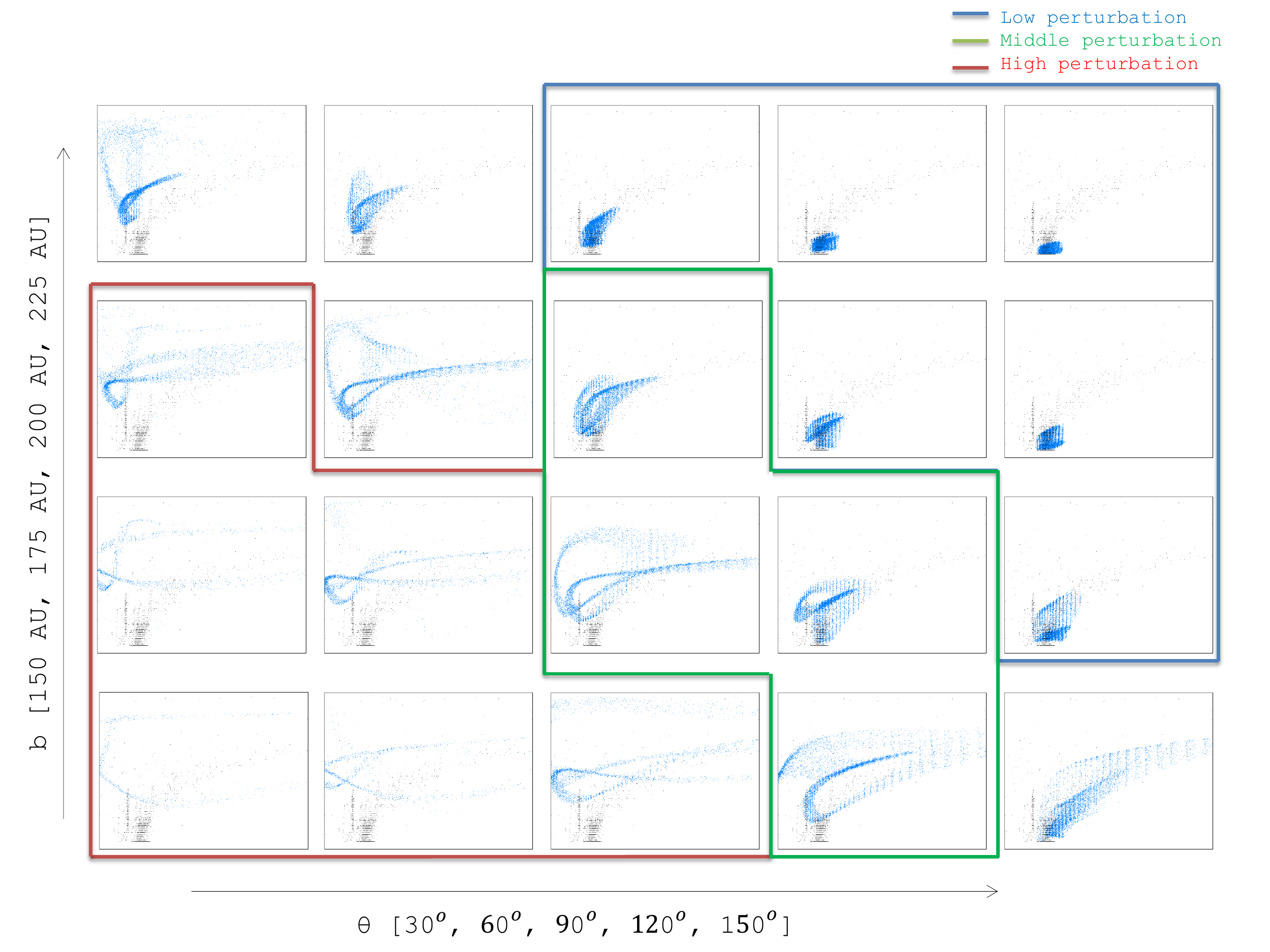}
\caption{Parameter coverage of the eccentricities (between 0 and 1)
  as a function of the semi-major axis (between 35 $\mathrm{AU}$ and 100 $\mathrm{AU}$) for
  model A at $10^{4}$, yr after the encounter. For readability we
  omitted the axis tickmarks and numbers, but each panel has identical
  axes as the images presented in the top panel of
  Fig.\,\ref{fig:oss}.  The mass of the passing-by star was
  1\,M$_\odot$ and the total mass of the Kuiper belt was assumed equal
  to $30$ M$_\oplus$. The black dots give the observed data from the MPC
  \citep{mpc} (see also Fig.\,\ref{fig:oss}) and the blue dots give the
  simulated data.  Each plot has different values of parameters $b$
  and $\theta$ ordered in increasing values. We identified three
  regimes in which the interaction had a strong influence on the
  distribution of KBOs (red), moderate (green) and mild (blue).  }
\label{fig:modA1}
\end{figure*}

\begin{figure*}
\centering
\includegraphics[width=1.0\textwidth]{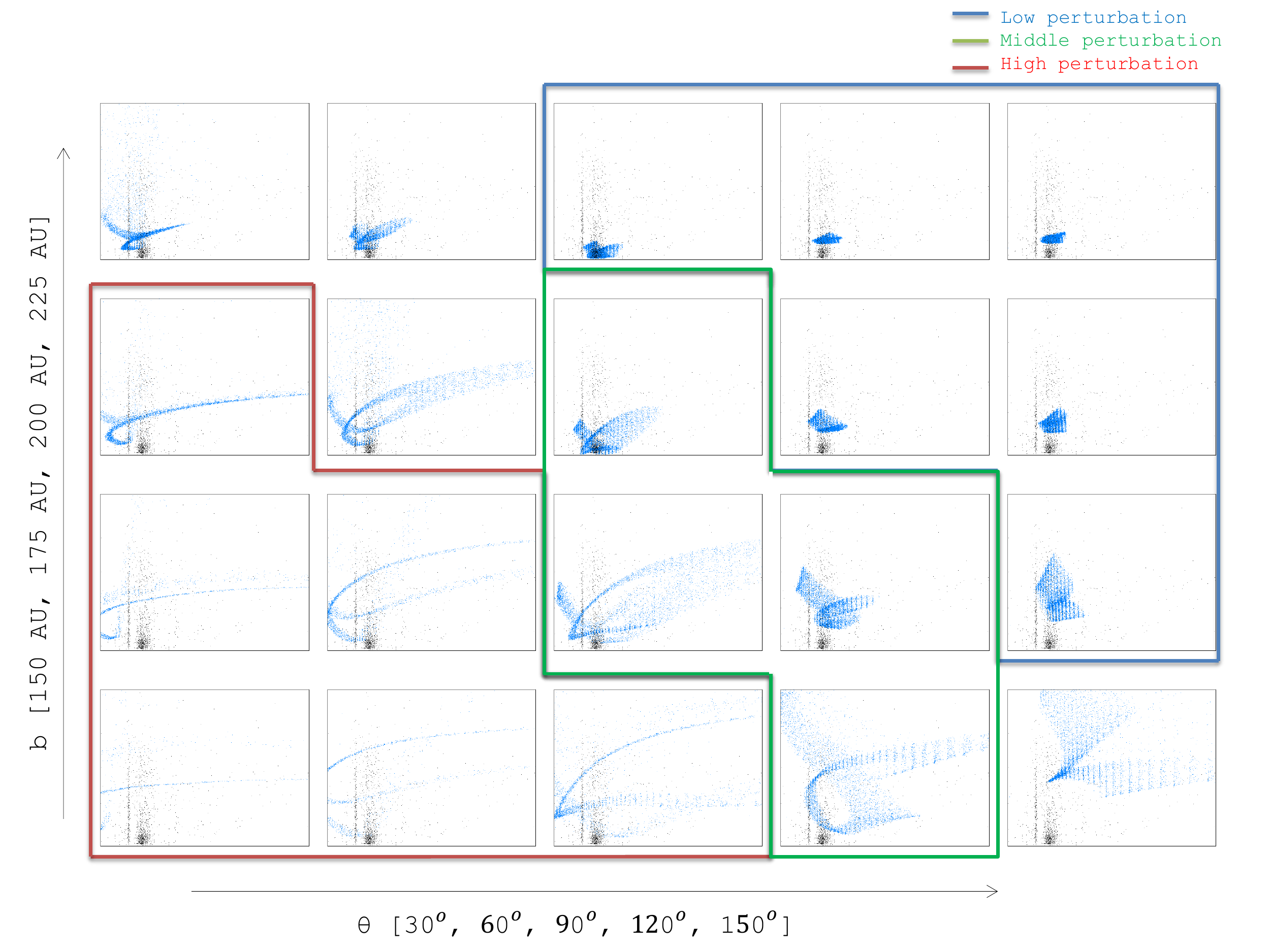}
\caption{As fig.\,\ref{fig:modA1} but then for the inclination
  (between 0 and $50^\circ$). }
\label{fig:modA2}
\end{figure*}

In Figs. \ref{fig:modA1} and \ref{fig:modA2} we show the distributions
for eccentricity and inclination as a function of the semi-major axis
for the KBOs from model A. Here we varied the impact parameter
parameters $b$ and inclination $\theta$ of the encounter. The mass of
the passing-by star was 1\,M$_\odot$ for these simulations and the
total mass of the KB was $30$ M$_\oplus$. These simulations were
performed with $N = 8182$ KBOs and run up to $10^4$\,yr. 

In the Figures we have identified three main regimes, which we
colored red, green and blue, indicating the highly, intermediate and
relatively little perturbed system, respectively.

In the highly perturbing encounter (red zone in figs.\,\ref{fig:modA1}
and \ref{fig:modA2}) the KB is almost completely destroyed.  In the
moderately perturbing encounter (blue zone) the post-encounter KBO is
characterized by that the majority of objects remain confined in the
classical region but with slightly elevated eccentricities and
inclinations. These distributions are most comparable to the observed
eccentricities and inclinations in the classic (observed) regions.
However, the resonance regions and the scattered region are
notoriously depleted compared to the observations.  In particular, the
distribution in inclinations is too much concentrated around a mean
value, whereas the observed inclinations are distributed more evenly
between $0^\circ$ to $40^\circ$.

The distribution of eccentricities in the mildly perturbed encounters
(green zone in figs.\,\ref{fig:modA1} and \ref{fig:modA2}) has almost
vanished and some objects have scattered to very small semi-major
axes. The general shape of the KBO, however, seems to follow the data
more closely that those that result from the other more strongly
perturbed interactions.  The majority of bodies resides in the
classical part of the KB with an extended tail of monotonically
increasing eccentricities with the semi-major axis, indicating an
almost constant periastron distance. On the down-side, however, the
distribution of inclinations is distinctively different than the
observations.

We compare the distributions in eccentricities and inclinations of the
KBOs at $10^4$\,yr after the encounter changing the total mass of the 
KB with values $1,3,6$ and $30$ M$_\oplus$  and the sampling of the 
KBOs with $N$ in range [8182, 131,062] fixing the $b = 200$ $\mathrm{AU}$ and 
$\theta = 90^\circ$ parameters.  This comparison is performed using 
the two-dimensional Kolmogorv-Smirnov tests \citep{recipes}.  The 
tests give probabilities for eccentricity as well as inclination $\geq
91.2\%$.  The K-S test is a measure of the difference in the two 
distributions. The high values of these K-S test is an indication that 
the distributions, obtained varying the principal parameters of the 
sampling of the KB, show very small differences. Therefore, a small 
variation of the initial conditions for the KB gives rise to only 
small changes in final distribution, and on the short time frame of 
the encounter, the effect of the passing star is considerably stronger 
than any internal dynamical effect inside the KB.  The effect the 
passing star has on the KB is almost impulsive, and variations in the 
mass of the passing star strongly affects the eccentricities and 
inclinations of the KBOs. These distributions therefore provide a 
sensitive characterization to constrain the mass and orbital 
parameters of the incoming star. 

It may be relevant noting that some of the consequences of the encounter
of star with the KB can be reliably predicted by the much simpler 
{\it test particle} approach, i.e. neglecting the internal interactions between 
the KBOs. Actually, a comparison of our results with test-particle
simulations \citep{Ida1,Ida2, Ida3, Melita}
show a certain level of similarities: 
\begin{enumerate}
\item a stellar encounter pumps up strongly the eccentricities and inclinations of 
objects in the outer region of a planetesimal disk. Moreover, if the classical KBOs 
acquire high eccentricities their perihelia migrate to the inside.
\item a strong stellar encounter (corresponding to star passing close to the KB disk)
may deplete the original, flat, KBO distribution up to 95$\%$. However, contrary to
\cite{Ida3}, we find that a strong depletion correspond to a full destruction of the
Solar system structure. On the other hand,  we found also reasonable initial
encounter conditions leading to ``intermediate" cases,  where a significant depletion (at about 
$13\%$ level) is compatible with the observed distributions of eccentricities and inclinations.
\item it is not possible to populate the observed resonances reproducing exactly the 
overdensity in the eccentricity and inclination distributions invoking only a fly-by 
star perturbation (see \cite{Ida1}). 
\end{enumerate}

The strong effect of the incoming star is clearly depicted in
Figs. \ref{fig:modA1} and \ref{fig:modA2}.  Varying the mass of the
encountering star cases a migration in both $b$ and $\theta$.  The
low-mass star (0.5\,M$_\odot$) gives rise to a shift to smaller values
of $b$ and $\theta$, whereas a higher mass star (2\,M$_\odot$) causes
a shift toward larger values of both parameters.

The early evolution of the system strongly depends on the initial
conditions of the passing star. We therefore decided to run more
simulations in the middle regime (green zone) in the range $\theta =
[70, 80, 90]^\circ$ and $b = [170, 180, 190, 200]$ $\mathrm{AU}$, and in a the
second regime of $\theta = [90, 100, 110, 120]^\circ$ and $b = [140, 
150, 160, 170]$ $\mathrm{AU}$. All the simulations show a characteristic tail to 
a monotonically increasing eccentricities; quite similar to the 
distribution of the eccentricities of the observed scattered KBOs.  
This tail is characteristic for the relatively close encounter with a 
stellar perturber, and we confirm the earlier made conjecture of such 
an encounter \citep{Ida1}.  The distribution in inclination, however, 
is still to easily reproduced. 

To validate our visual comparison we performed a statistical
cross-comparison between the observational and computational data for 
each of the simulations in Tab.\,\ref{tab:tabcond} using the 
Hotelling's two sample $T^2$ test \citep{Hotelling},
which is a generalization of the Student's $t$ test,  where

\begin{equation}
F = \frac{n_x + n_y - p - 1}{(n_x + n_y -2)p} T^2,
\end{equation}  
with $F$ the Fisher-Snededecor random variable and $T^2$ is defined as:
\begin{equation}
T^2 = \left(\bar{X}-\bar{Y}\right)^T \left[S \left(S\frac{1}{n_x}+\frac{1}
{n_y}\right)\right]\left(\bar{X}-\bar{Y}\right),
\end{equation}  
where $S$ is the pooled sample covariance matrix of $X$ and $Y$, namely, 
\begin{equation}
S = \frac{(n_x -1)S_x + (n_y -1)S_y}{(n_x -1) + (n_y -1)}
\end{equation}  
where $S_x$ is the covariance matrix of the sample for $X$, $\bar{X}$ is 
the mean of the sample, and the sample for each random variable $x_i$ in 
$X$ has $n_x$ elements, and similarly $S_y$ is the covariance matrix of 
the sample for $Y$, $\bar{Y}$ is the mean of the sample, and the sample 
for each random variable $y_i$ in $Y$ has $n_y$ elements.
The Hotelling's test states that two population are indistinguishable if
\begin{equation}
F \lesssim F_{tab}(p, n_x + n_y - 1 - p, \alpha),
\end{equation}  
where $p$ is the number of parameters, $\alpha$ the significance
level of the test and $F_{tab}$ is the theoretical value of $F$-distribution.
In Tab. \ref{tab:tabA} we present the values for $F$. Each $F$-value is the 
comparison between two samples: the observed KBOs and the computational one.
In order to calculate the $T^2$ variable we compute it using the JD2000 Ephemeris 
(Right Ascension and Declination coordinates) of the 
observed KBOs and on the coordinates of the computational KBOs converted in 
equatorial coordinates. 
We have normalized the values with $F_{tab}(2, N + n - 3, 0.10) = 9.49122$
(where $n_x$ is $N$ our number of KBOs parameter and $n_y$ is $n = 
1593$ the number of observed KBOs \citep{mpc}).  In the table (Tab. 
\ref{tab:tabA}) we have subtracted one from the result to 
make the clearer distinction that negative values represent results 
for which the computational and the observational distributions are 
statistically indistinguishable.
We did not perform the Hotelling's test directly on the eccentricity 
and inclination distributions because the test can be calculated 
only on two samples that have a \textit{normal} distributions. 
On the other hand, ($\alpha$,$\delta$) are not independent from ($e$,$i$) 
values, therefore a negative value in Tab. \ref{tab:tabA} tells also 
information about the eccentricity and inclination distributions.

This test suggests that encounters with a large impact parameter $b$ 
or large inclination $\theta$ are favored (strictly speaking, the 
other runs are rejected on this statistical).  Although this method 
does not makes a distinction in quality of the results, other than
accepting or rejecting it, the area of parameter space that give
small perturbations (blue zone) seems to be favored.  These 
simulations show a cold and low-eccentricity population of KBOs.

\begin{figure}
\centering
\includegraphics[scale=3.00]{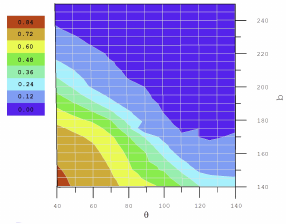}
\caption{The 2-D surface distribution of the fraction of the KBO 
escaped after the gravitational encounter for the model A, at $t = 
10^4$ $yr$, in function of the parameters $b$ and $\theta$. In the 
color map is reported for each color the mean percentage of escaper 
for that zone in the parameter space.}
\label{fig:escaper}
\end{figure}

In Fig. \ref{fig:escaper} we present the fraction of KBOs that escaped
the Solar System as a result of the stellar encounter.  Based on these
results we prefer the highly scattered regime, with a low value for
$b$ and $\theta$ (red zone). In fact only in the highly perturbed 
regime there is a substantial loss of mass which can reconcile the 
difference of two magnitude between the total mass observed and the 
total mass predicted by Solar System formation models \citep{Luu}. 

These contradicting results let us to perform a second series of 
simulations in which we adopted a wider range of semi-major axis, this 
we called model B.

\begin{table}
\centering
\begin{tabular}{c|ccccccccc}
\hline 
\hline 
\LARGE{$_b \backslash ^{\theta}$} & 30 & 60 & 70 & 80 & 90 & 100 & 110 & 120 & 150\tabularnewline
\hline 
225 & -0.1 & -0.7 & - & - & -0.8 & - & - & -0.9 & -0.8\tabularnewline
200 & 7.0 & 1.3 & 0.8 & 0.2 & 0.0 & - & - & -0.7 & -0.6\tabularnewline
190 & - & - & 1.6 & 0.9 & 0.1 & - & - & - & -\tabularnewline
180 & - & - & 2.0 & 1.3 & 0.8 & - & - & - & -\tabularnewline
175 & 4.7 & 2.3 & 1.3 & 0.3 & 0.4 & - & - & -0.6 & 0.0\tabularnewline
170 & - & - & - & - & 0.5 & 0.7 & -0.4 & - & -\tabularnewline
160 & - & - & - & - & -0.5 & 0.1 & -0.2 & - & -\tabularnewline
150 & 2.3 & 1.1 & - & - & -0.1 & 0.1 & -0.1 & -0.3 & 0.1\tabularnewline
140 & - & - & - & - & -0.5 & 0.0 & 0.5 & - & -\tabularnewline
\hline
\hline
\end{tabular}\caption{Values of the $F$ indicator obtained with 
different parameters $b$ (listed in the left column) and $\theta$ (in 
the upper row) for the model A.}
\label{tab:tabA}
\end{table}

\subsection{Model B}

In model B we adopted a wider radial extension of the KBO, 42 $\mathrm{AU}$ to
90 $\mathrm{AU}$ with a total mass of $M = 30$\,M$_\oplus$.  The mass and
encounter parameters of the incoming star are presented in
Tab.\,\ref{tab:tabcond}.  The success of model A in reproducing the
observed parameters for the KBO led us to limit our parameter search
this more extended distribution to $b = [200;237.5]$ and $\theta =
[60^\circ;150^\circ]$. The results of the Hotelling test are
presented in Tab.\,\ref{tab:tabB}.

\begin{table}
\centering
\begin{tabular}{c|ccccccc}
\hline 
\hline 
\LARGE{$_b \backslash ^{\theta}$}  & 60 & 75 & 90 & 105 & 120 & 135 & 150\tabularnewline
\hline 
237.5 & -0.6 & -0.8 & -0.8 & -0.9 & -0.8 & -0.4 & -0.7\tabularnewline
225 & -0.2 & -0.8 & -0.6 & -0.7 & -0.9 & -0.2 & -0.7\tabularnewline
212.5 & -0.3 & -0.5 & -0.7 & -0.7 & -0.8 & -0.9 & -0.4\tabularnewline
200 & 0.1 & -0.3 & -0.7 & -0.8 & -0.5 & -0.9 & -0.4\tabularnewline
\hline 
\hline 
\end{tabular}
\caption{ Values of the $F$ indicator for various values of the
  parameters $b$ (left column) and $\theta$ (upper row) for the model
  B.}
\label{tab:tabB}
\end{table}

\begin{figure}
\centering
\subfigure{
\includegraphics[width=0.5\textwidth]{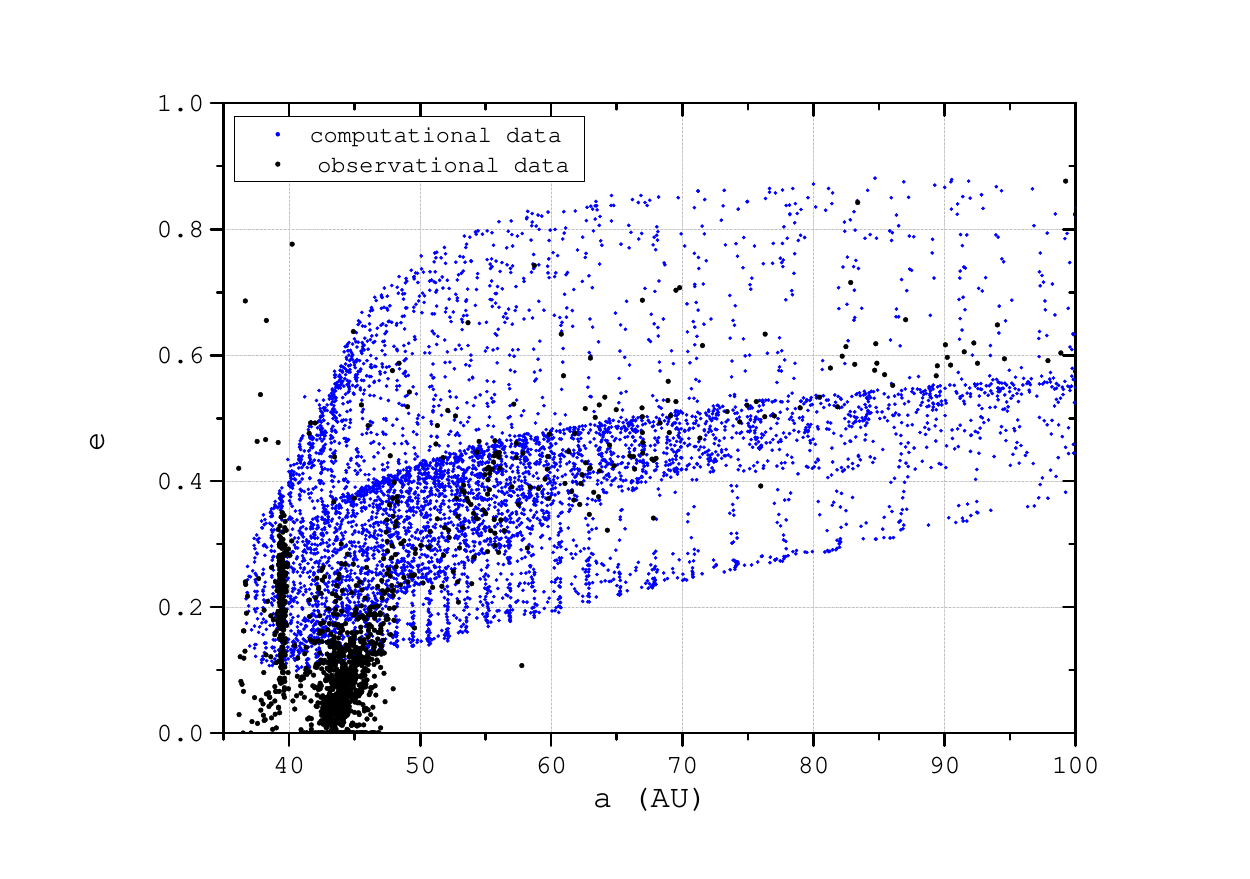}}
\subfigure{
\includegraphics[width=0.5\textwidth]{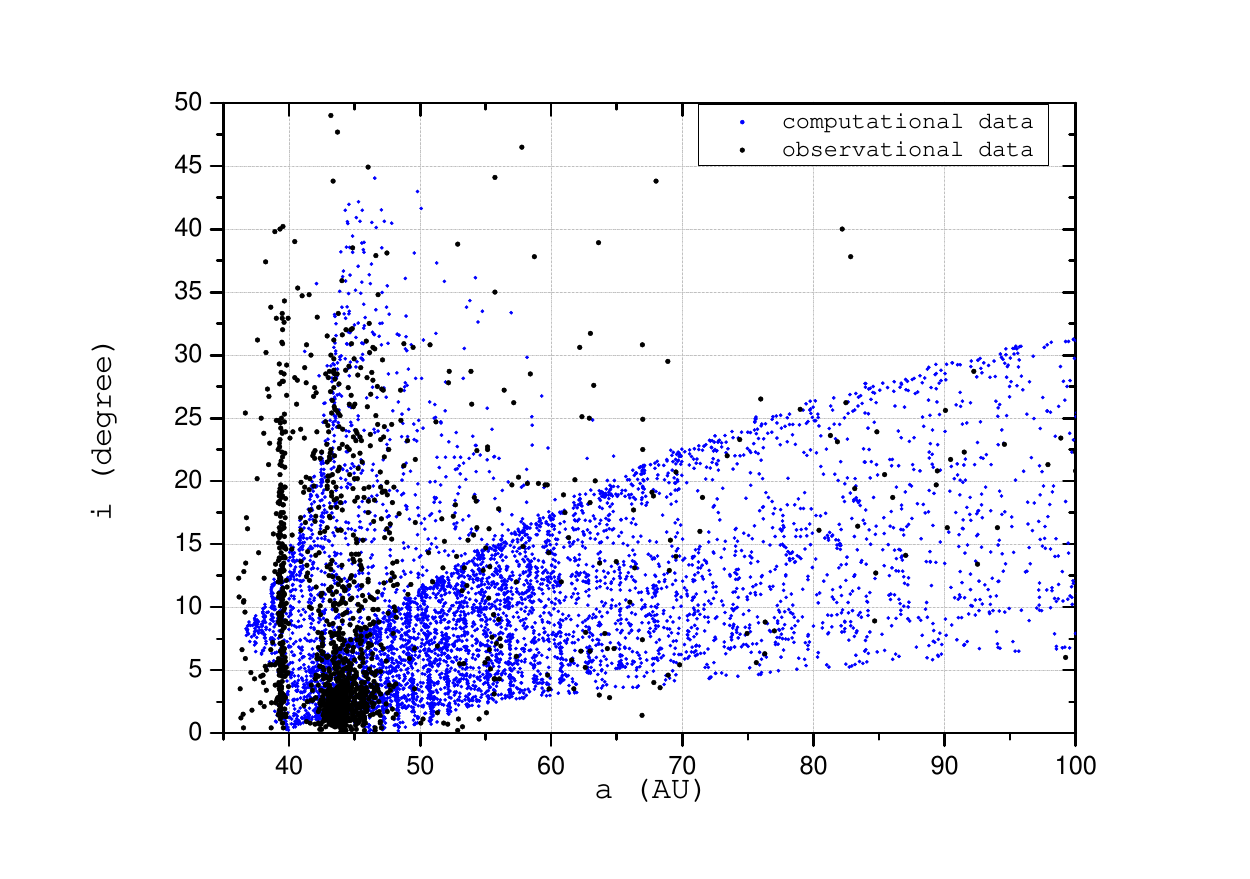}}
\caption{Comparison between the observe parameters and the result of
  model B directly after the scattering (at $T = 10^4$\,yr) of an
  encounter with $b = 200$ $\mathrm{AU}$, $\theta = 90^{\circ}$.  The top panel
  gives the eccentricity vs. the semi-major axis, the lower panel is
  for the inclination.  }
\label{fig:modelB}
\end{figure}

In Fig. \ref{fig:modelB} we present the distributions for eccentricity
and inclination for $b = 200$ $\mathrm{AU}$ and $\theta = 90^\circ$.  We 
determine it as the best model which gives a much better match with 
the observed inclinations; with value ranging from roughly $0^\circ$ 
up to $40^\circ$ in the classical regime and up to $30^\circ$ in the
scattered regime whereas the eccentricities of the scattered
population and the high-eccentricity population 
in the classical region are consistent with the observational data.  
With these parameters the initial KB lost $\sim 13\%$ of its mass in 
the encounter, which is quite small compared to the predictions \citep{Luu}.

The parameters for this particular encounter has trouble reproducing
the low-eccentricity population; in fact, the 
minimum value of the eccentricities in the classical regime $~0.1$.
For this reason we started a series of simulations in which we study 
the long-term secular evolution of the KB, on which we
report in \S\,\ref{secularevo}.

\begin{figure}
\centering
\subfigure{
\includegraphics[width=0.5\textwidth]{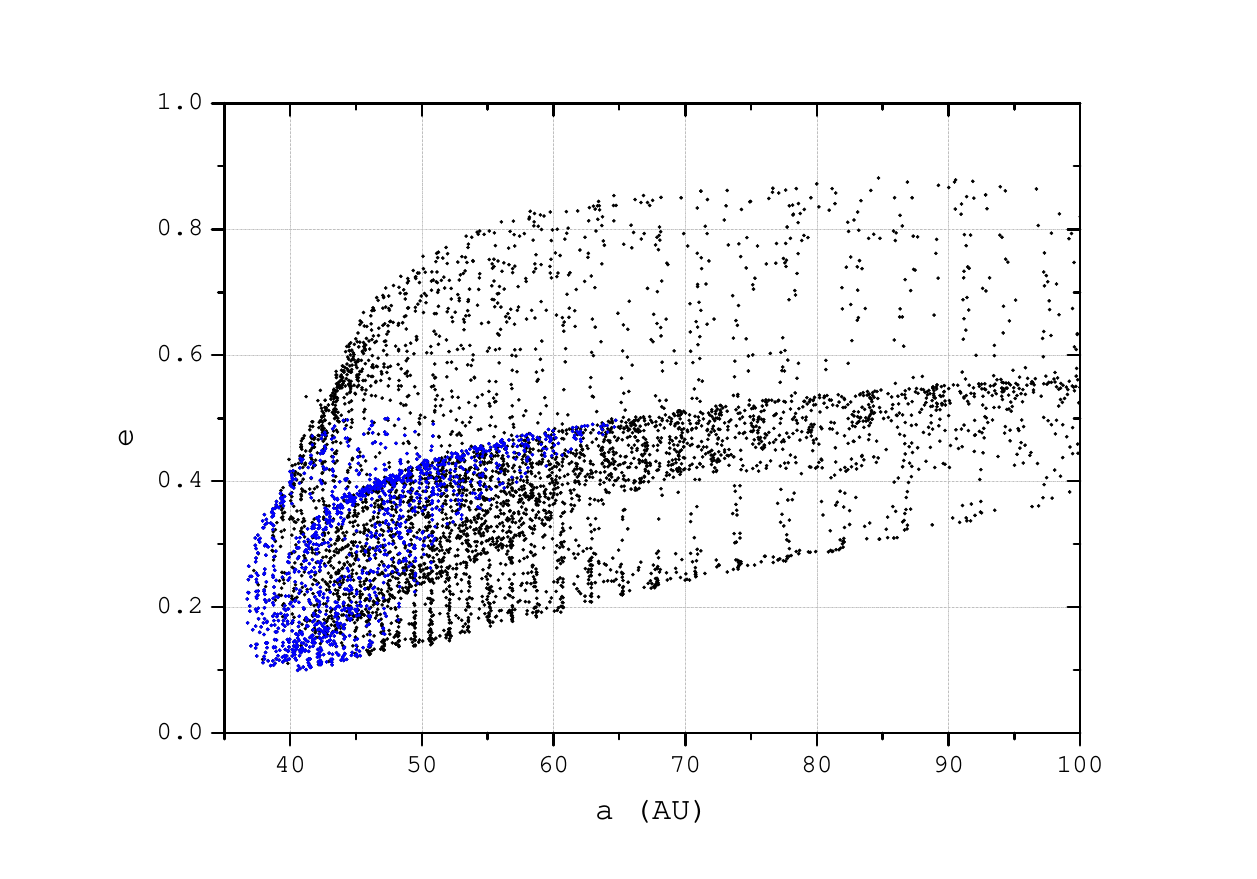}}

\subfigure{
\includegraphics[width=0.5\textwidth]{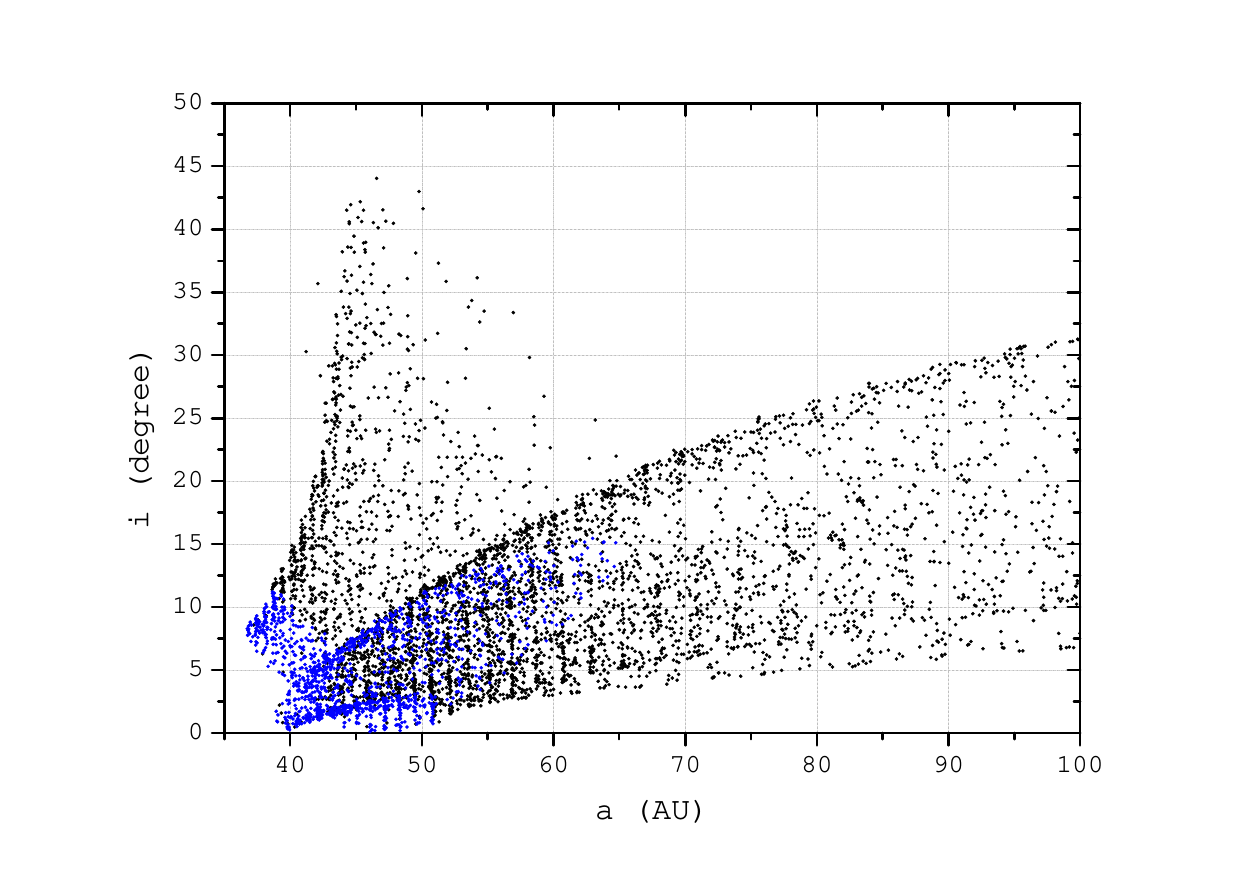}}
\caption{Distribution of eccentricities (top panel) and inclinations
  (bottom panel) of model A (blue) with model B (black) directly after
  the encounter (at $T = 10^4$\,yr) with parameters $b = 200$ $\mathrm{AU}$ and
  $\theta = 90^{\circ}$. This comparison is to demonstrate that model
  A is indeed a subset of model B, but with a higher resolution at the
  short-period KBOs. }
\label{fig:modelBA}
\end{figure}

For comparison we highlighted, in Fig. \ref{fig:modelBA} the
eccentricities after the encounter for model A and model B for one
particular encounter. For the range where the initial conditions of
model A and B overlap, the post-encounter distributions in 
eccentricity and inclination also overlap. This supports the earlier 
argument that inter-KBO dynamics is not important during the 
encounter.

\subsection{Long term evolution}\label{secularevo}

The main result for model A and B has been that an encounter in the
early history of the Solar System can reproduce the high-eccentricity
KB population as well as the majority of the scattered population, but 
the currently observed low-eccentricity population 
and part of the resonant populations are absent after the encounter. 
We will now investigate if these missing populations can be regrown 
by the long-term evolution of the KB.

We adopt model A with a 1\,\MSun\, encountering star with an impact
parameter $b = 200$ $\mathrm{AU}$ and inclination $\theta = 90^\circ$ for
this follow-up study.  Ideally, we should have taken the best model B,
but because the missing populations are reachable with the limited
range in semi-major axes in model A, we decided that the benefit of
the higher local resolution of this model outweighs the more extended
width of model B.  We restart the simulation at $10^4$\,yr after the
encounter. For convenience we removed the encountering star, because
it would cause numerical problems if we allowed it to continue to move
further away from the Solar System, whereas it would no longer perturb
the KB.   

\begin{figure*}
\subfigure{
a) \includegraphics[width=0.4\textwidth]{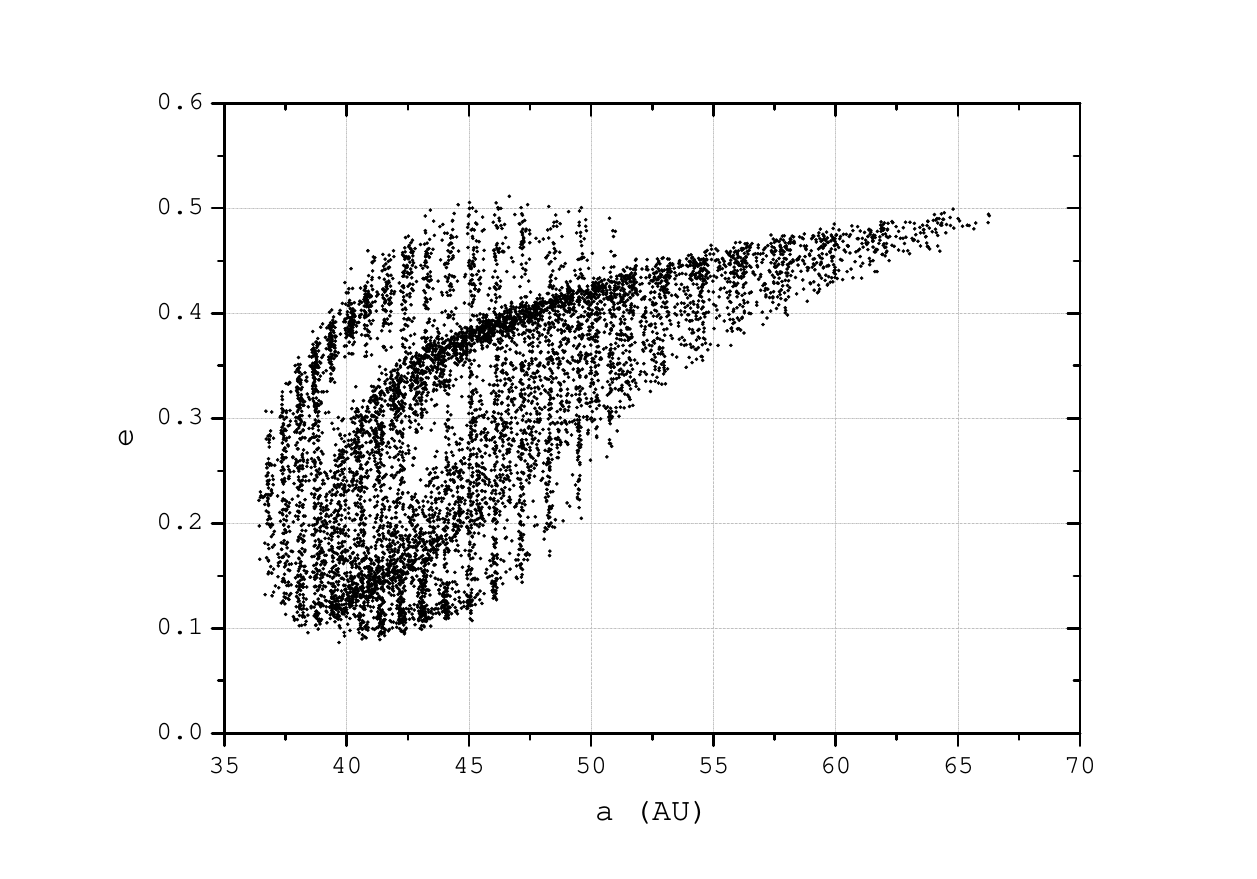}}
\subfigure{
b) \includegraphics[width=0.4\textwidth]{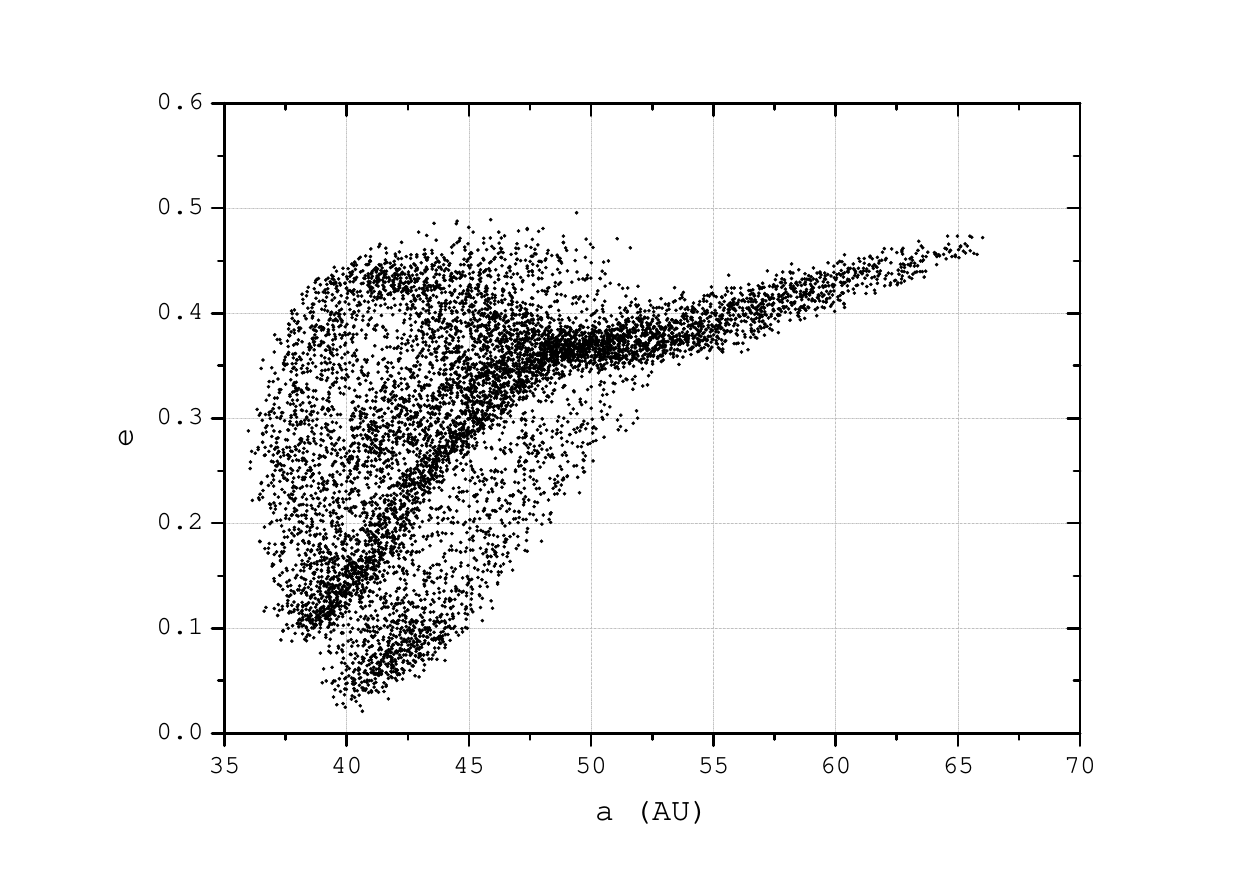}}
\subfigure{
c) \includegraphics[width=0.4\textwidth]{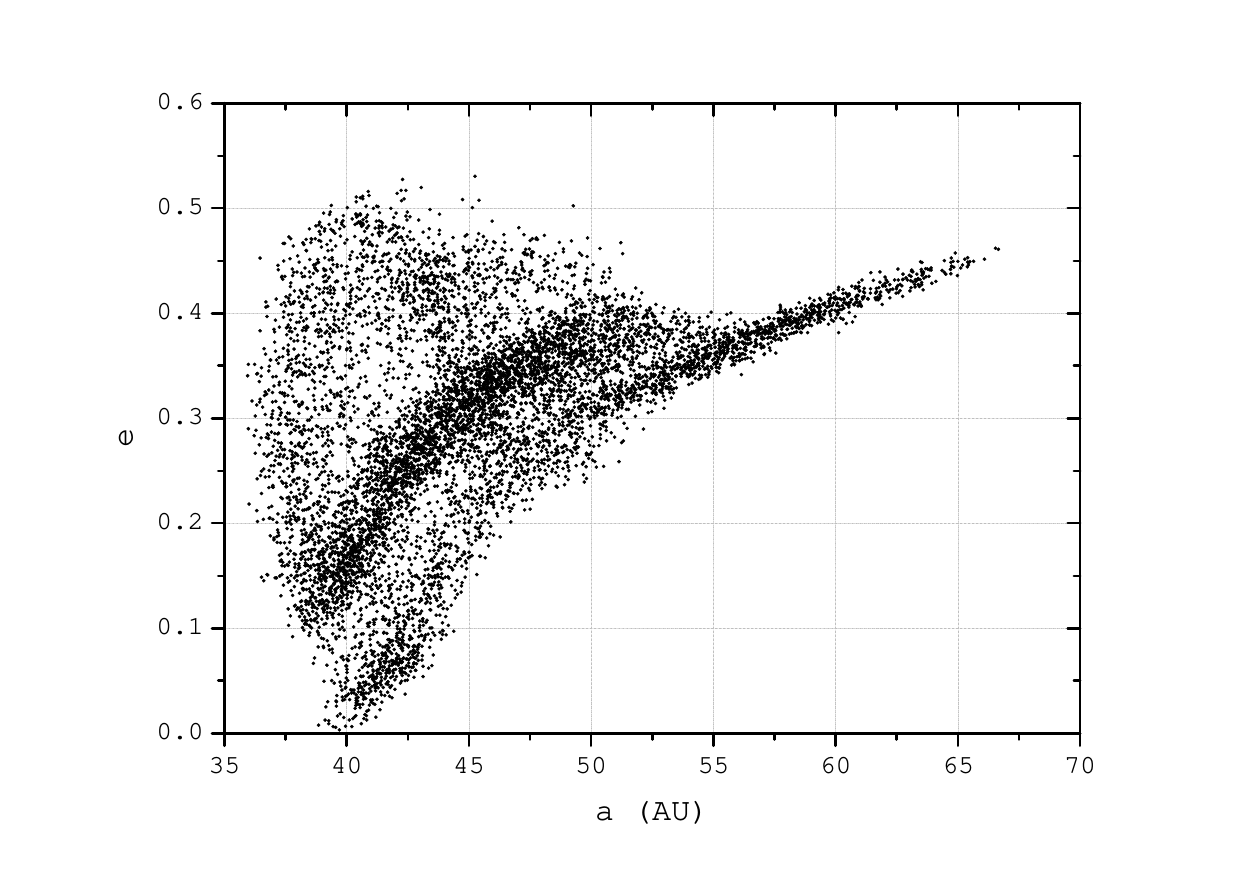}}
\subfigure{
d) \includegraphics[width=0.4\textwidth]{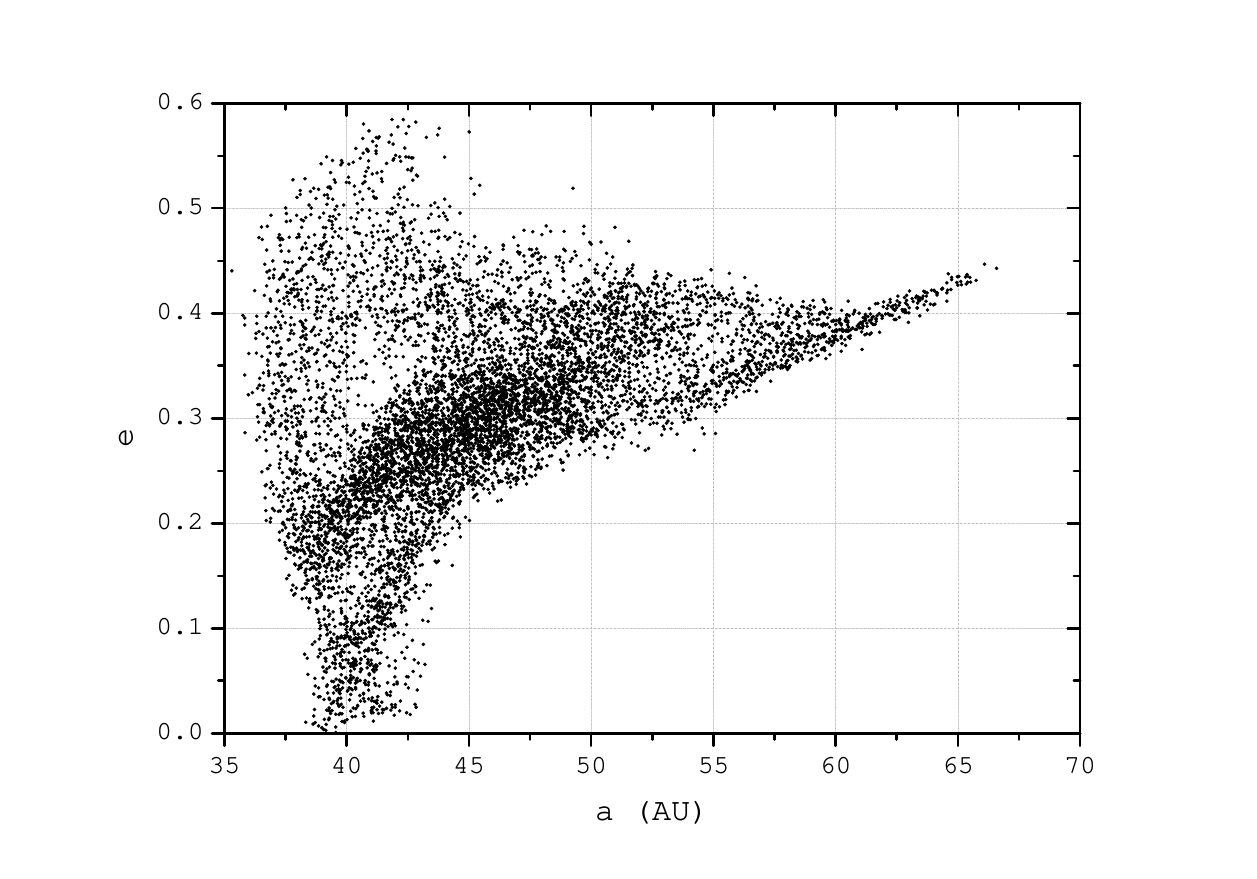}}

\caption{Time evolution of the distribution of KBOs (model A) due to
  the internal dynamical relaxation. The encountering star was
  1\,\MSun\, , with impact parameter $b = 200$ $\mathrm{AU}$ and inclination 
   $\theta = 90^{\circ} $.  The snapshots were taken at a) $10^4$ yr,
  b) $1.6 \times 10^5$ yr, c) $3.2 \times 10^5$ yr and frame d at $0.5
  \times 10^6$ yr.}
\label{fig:secular1}
\end{figure*}

In Fig.~\ref{fig:secular1} we present the evolution of the
eccentricities and semi-major at four moments in time. This 
illustrates the effect of the secular evolution of the KBOs, due to 
their self gravity and the influence of the planets.  Whereas the 
encounter with the star completely removes the 
low-eccentricity population, 
the subsequent long-term evolution within the KB regrows this 
population.  During the secular evolution the high eccentricity orbits
contained between $37$ $\mathrm{AU}$ and $46$ $\mathrm{AU}$ ``cool" to
lower eccentricities.  In this model we considered $8182$ KBOs 
($N=8182$) with a total mass of 30 M$_{\oplus}$; each KBO, then, has the 
mass $3.7\times 10^{-3}$ M$_\oplus$.  

An important test of the importance of mutual gravitational interactions
between KBOs in determining the KB secular evolution has been done 
through the expedient of ``switching off'' the pair interaction. We saw
that these simulations did not show the ``cooling" of the KB populations,
with eccentricities which remained too high compared to the observations.
We were driven to conclude that the mutual interactions among the KBOs are
responsible of the secular evolution to the partial repopulation of 
the low-eccentricity distribution  after a stellar encounter.

In Tab. \ref{tab:sec} we 
present a summary of results of several simulations at varying $N$ and the 
KBO radius, which correspond to a variation of the individual KBO mass. 
Initial conditions are those of model A. We noted that, 
as expected, the time needed to  repopulate the \textcolor{blue}{low-eccentricity} 
KBO distribution ($T$ in the right most column of Tab. \ref{tab:sec}) scales 
roughly as the two body relaxation time scale, which, in a virialized system, 
has the following dependence on $N$ and $m$ \citep{binney}:

\begin{equation}
t_{rel} \propto \frac{N}{ln(N)} \frac{1}{\sqrt{N\,m}}. 
\label{eq:rel}
\end{equation}

\begin{table}
\centering
\begin{tabular}{|c|c|c|c|c|}
\hline 
label & $N$ & $m\:($M$_\oplus)$ & $R$ (km)& $T$ (Myr)\tabularnewline
\hline 
1 & $8182$ & $3.7\times10^{-3}$ & 1741 & 0.50\tabularnewline
\hline 
2 & $8182$ & $3.7\times10^{-4}$ & 808 & 1.58\tabularnewline
\hline 
3 & $8182$ & $3.7\times10^{-5}$ & 375 & 5.00\tabularnewline
\hline 
4 & $4086$ & $3.7\times10^{-3}$ & 1741 & 0.77\tabularnewline
\hline 
5 & $65526$ & $3.7\times10^{-3}$ & 1741 & 1.15\tabularnewline
\hline 
6 & $65526$ & $3.7\times10^{-4}$ & 808 & 3.64\tabularnewline
\hline 
7 & $81820$ & $7.0\times10^{-7}$ & 100 & 91.56\tabularnewline
\hline
\end{tabular}\caption{Entries are: $N$, the number of KBOs;  
$m$, the mass of the single KBO, in units of Earth mass; 
$R$, the radius 
of the single KBO; $T$, the time-scale to repopulate 
the \textcolor{blue}{low-eccentricity} KBO distribution. 
Actually, for the simulations $\#$ 5-6-7 we checked only the initial cooling phase and 
then the actual values of the parameter $T$ were extrapolated using Eq.$\,$\ref{eq:rel}. }
\label{tab:sec}
\end{table}

After establishing the initial conditions which we considered to
produce the observed KB, we run one more simulations, with $N=16,374$ 
and a total mass of 30\,M$_{\oplus}$ (35\% of the KBOs have a mass 
identical to Pluto, while the others have only one-fifth of this 
mass). The incoming 1\,\MSun\, star has approaches the Sun with impact 
parameter $b = 200$ $\mathrm{AU}$  and inclination $\theta = 90^\circ$. We 
continue this simulation for 0.9\,Myr.

\begin{figure}
\centering
\subfigure{
\includegraphics[width=0.5\textwidth]{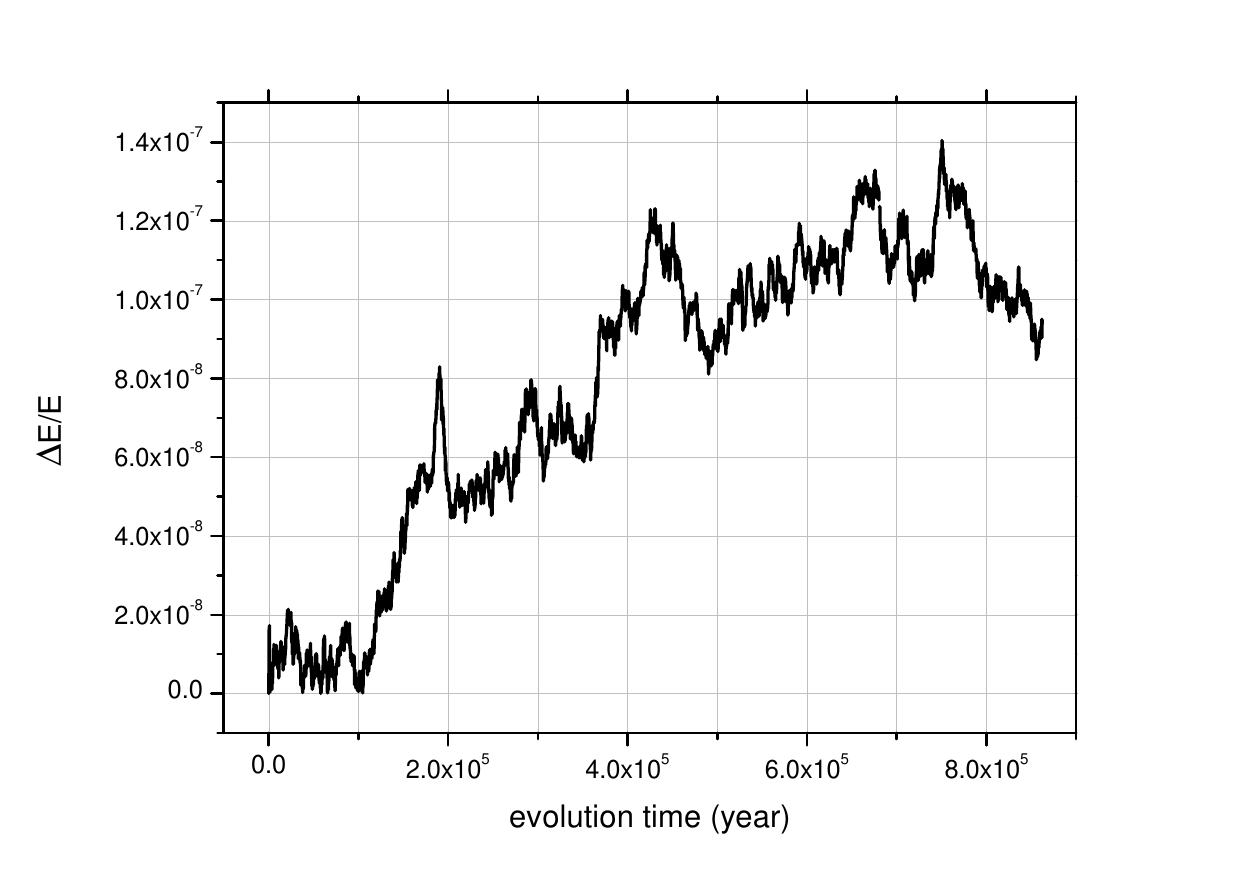}}
\subfigure{
\includegraphics[width=0.5\textwidth]{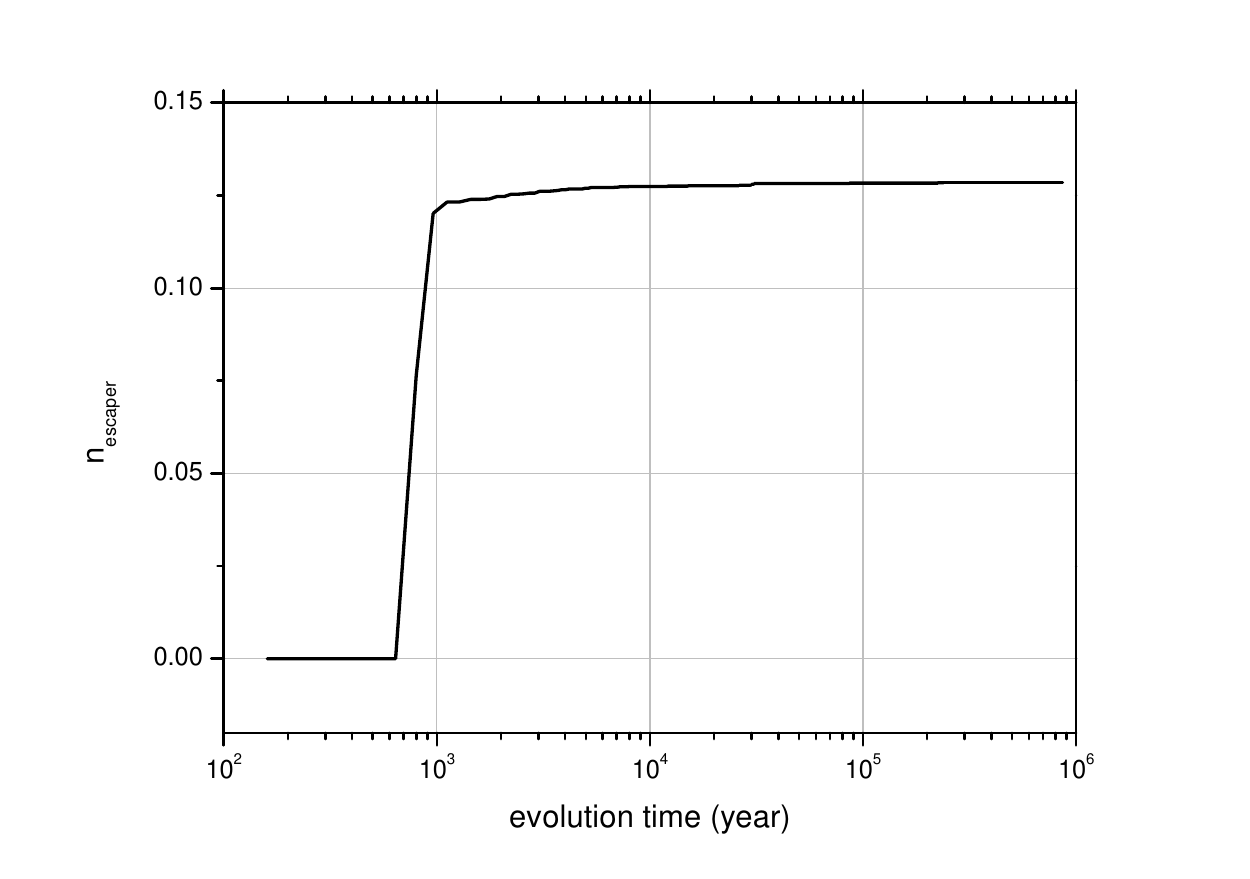}}
\caption{The relative error in the energy ($|\Delta E/E|$) (top panel) 
and the fraction of escapers (bottom panel) as functions of 
time for a 1\,\MSun\, encounter with $b = 200$ $\mathrm{AU}$ and $\theta =  90^\circ$. }
\label{fig:secular2}
\end{figure}

\begin{figure}
\subfigure{
\includegraphics[width=0.5\textwidth]{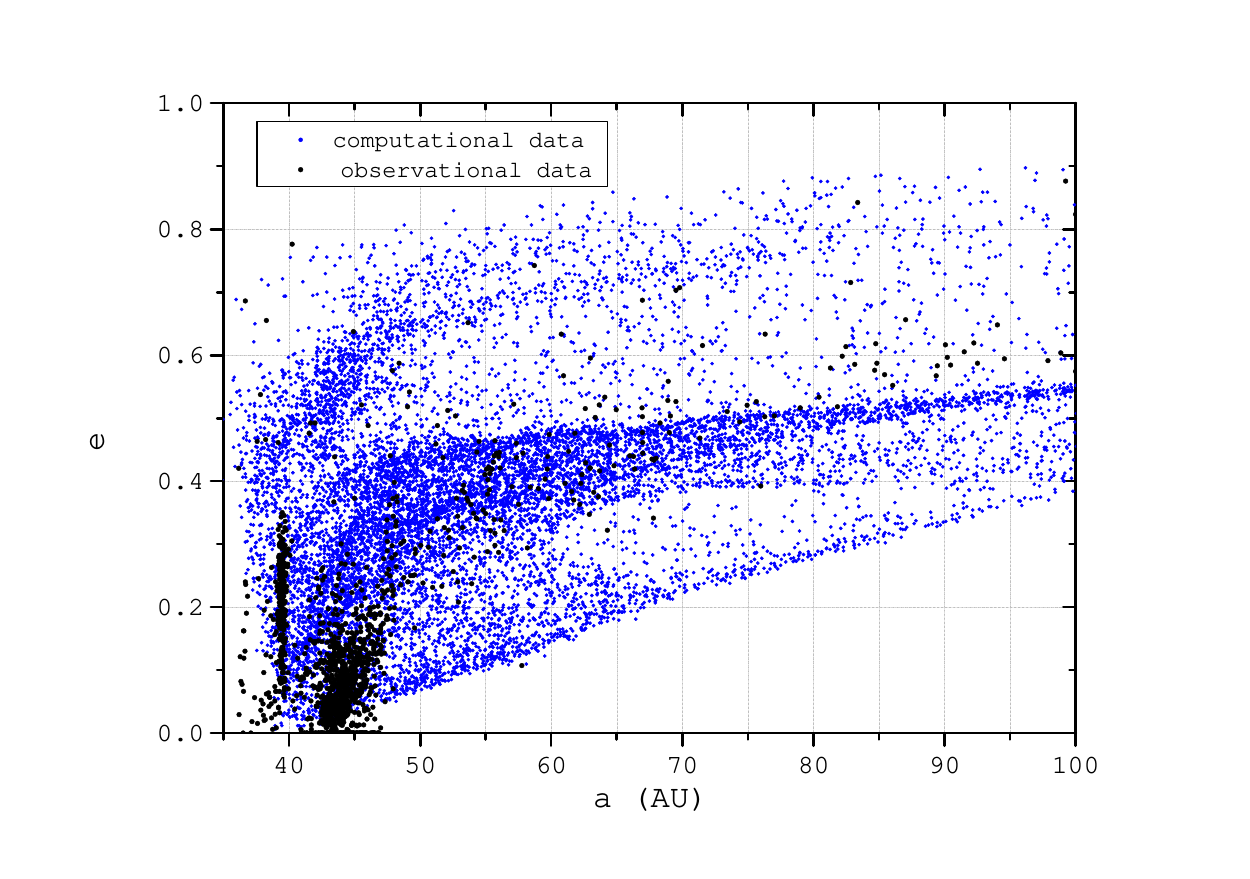}}
\caption{The eccentricities of the model B with encounter parameters
  $b = 200$ $\mathrm{AU}$ and $\theta = 90^{\circ}$ at $0.9 Myr$. The black dots
  are the observed data from the MPC \citep{mpc} and the blue ones are
  the simulated data using our best model.}
\label{fig:secular3}
\end{figure}

In Fig. \ref{fig:secular2} we present the energy conservation and the
total number of escapers in function of time. The distribution of the
eccentricities at 0.9\,Myr after the encounter is shown in
Fig.\ref{fig:secular3}.

\section{Conclusions}

We investigated the effect of an encounter between a passing star on
the morphology of the Kuiper belt, and its subsequent long-term
evolution. Using the current morphology of the KB we constrained the
parameter of the incoming star.  The orbit of the encountering star,
the planets and those of the KBOs were integrated directly, as was the
subsequent evolution of the internal dynamics of the KB and planets.
The initial conditions for the Solar System ware taken from
\cite{Ito}, and the Kuiper belt objects were distributed in a flat
disk between 42 and 90 $\mathrm{AU}$ in the plane of the Ecliptic, and with a
power law density distribution with exponent $-2$. The total mass of
Kuiper belt ranged between 1 and 30 M$_\oplus$, and the mass of the
incoming star was chosen to be 0.5\,\MSun\, 1.0 and 2.0\,\MSun.

\begin{figure}
\centering
\includegraphics[width=0.5\textwidth]{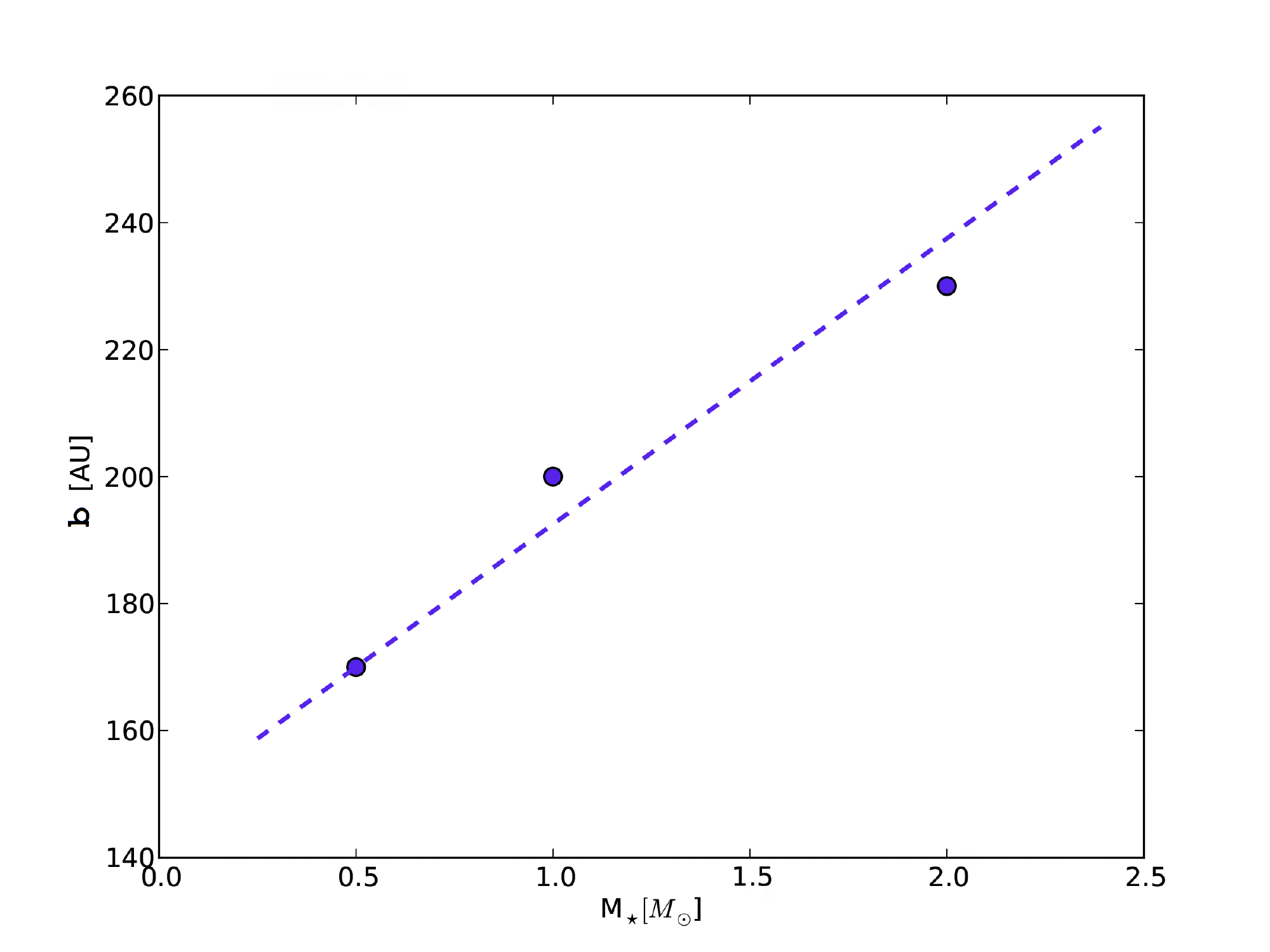}
\includegraphics[width=0.5\textwidth]{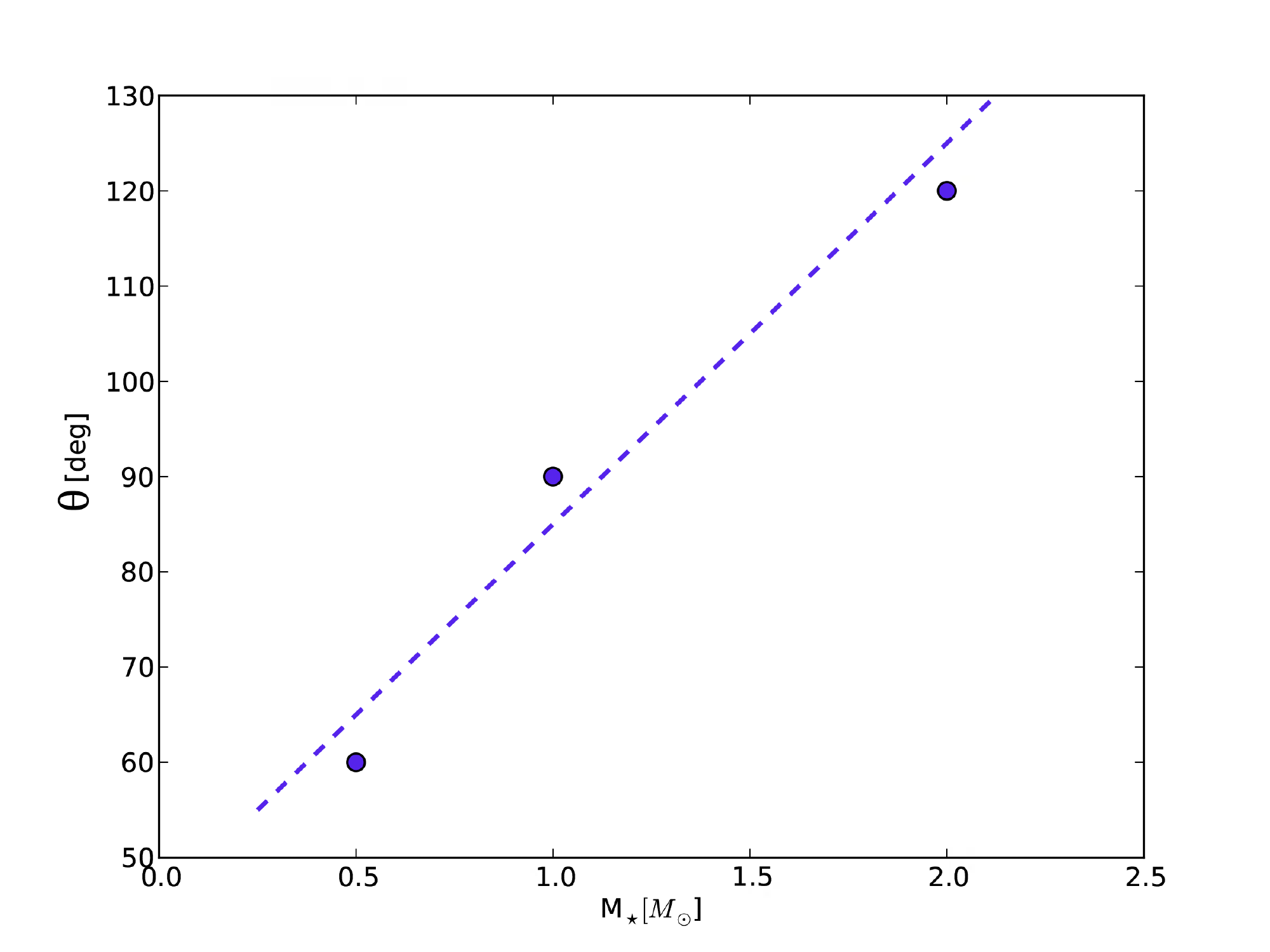}
\caption{The impact parameter $b$ (top) and inclination $\theta$
  (bottom) as a function of the mass of the incoming star for our most
  favorite model (see Tab.  \ref{tab:conc}). The dashed line gives a
  fit to the tree points to indicate the trend, which follows $b = 170
  + 45(M-0.5)$ for the impact parameter and $\theta = 65 +
  40(M_\star-0.5)$ for the inclination.}
\label{fig:Mstar_vs_d}
\end{figure}

We compared the morphology of the KB directly after the encounter with
the passing star, and after a secular evolution of up to 0.9\,Myr.
The best results, directly after the encounter, are obtained when the 
incoming star approached the ecliptic plane with an impact parameter of 
170-220 $\mathrm{AU}$ and an inclination above the Ecliptic of $60^\circ$ to 
$120^\circ$. The lower (best) values of both $b$ and $\theta$ are for
the 0.5\,\MSun, encountering star whereas the upper values correspond 
to the 2.0\,\MSun\, intruder.  We summarize these 
results in Tab. \ref{tab:conc}. In Fig. \ref{fig:Mstar_vs_d} we present the impact 
parameter and the angle $\theta$ of the incoming star as a function of its mass. 
A correlation between these parameters is evident and shows a degeneration in the 
parameters space. In fact, using different parameters is possible to reproduce
an encounter with the same strength and find similar proprieties in the 
final KBOs distributions.

During this encounter about 13\% of 
the Kuiper belt is lost from the Solar system. Actually, results do not 
show a depletion of the original flat distribution up to $\sim 99$ $\%$ as
suggested by the observed total mass of the KB, evaluated in the range
$0.3-0.1$ M$_\oplus$, and the mass estimation from Solar system formation model,
$30-10$ M$_\oplus$, \citep{Luu} and match the eccentricity and inclination 
distributions with the observation at the same time. On the other hand, a better 
coverage of the initial conditions of the incoming star can very likely enhance the
possibility of finding an {\it intermediate} case,  where a strong depletion can be
compatible with the observed distributions in eccentricities and inclinations.

The morphology of the high-eccentricity and scattered 
population of the KB are well represented directly after the encounter.  
The low-eccentricity population, around $\sim 40$ $\mathrm{AU}$
and with eccentricities $\aplt 0.1$ is almost completely absent directly
after the encounter. This mismatch in the morphology can be resolved by
taking the secular evolution of the Kuiper belt into account. 
The low-eccentricity population is 
reinstated within a million years. 
Our models did not show any particular correlation between the inclinations 
distributions and the mass of the KBOs. However, due to the limited number of
objects in our simulations, we could run only almost single-mass particle simulations. 
For example in our best model the gap in mass between the two population is only 
a factor 5 and the ratio between the radius is 1.7. 
Due to this limitation it is not possible to constrain 
any significant correlations among inclination and other 
properties, such as the size distribution of the KBOs and the number 
of KBO binaries. In conclusion, the sampling limiting our model and the 
relatively short time-scale of our simulations cannot give reliable results 
on that (actually, our finest simulation involved 16384 KBOs and was carried 
up to 0.9 Myear).

We expect that a more sophisticated investigation of the long-term ($\sim 100\,$Myr) 
evolution of the KB, with a proper population over the whole KBO mass spectrum  
will show the ``relaxation" of the eccentricities to low values as it happens 
in  the case of mass monodisperse-particle simulations. Such detailed studies 
would thus provide important information about the final distribution of the 
KBOs, which will allow a complete comparison with observable such as the 
size-inclination relation, but unfortunately it is hard to achieve at the 
moment without the access to a very large GPUs cluster.

While the secular evolution repopulated the low-eccentricity 
population, it triggered the further KB causing the 
depletion of the resonance population, which was initiated by the passing star.  
This loss of the resonant population can be due to the insufficient sampling
of the KB in our simulations. Alternatively, the early migration of the
planets is driving the repopulation of the resonant families
\citep{Mal2, Ida1}.  Such planetary reordering would be a natural
consequence of the Nice model \citep{Levi}. Moreover, the resonances, 
that we have suddenly after the passage of the fly-by star, do not show an 
eccentricity and inclination distributions compatible with the observations.

\begin{table}
\centering
\begin{tabular}{|c|c|c|c|}
\hline
M  & 0.5 & 1.0 & 2.0 \tabularnewline
\hline
$b$        & $170$ & $200$ & $220$               \tabularnewline
$v_{\infty}$& $3$   &  $3$   & $3$               \tabularnewline
$\theta$   & $60$ & $90$  & $120$                \tabularnewline
\hline
\hline 
\end{tabular}
\caption{The optimal encounter parameters ($b$ and $\theta$) obtained for a star with mass 
$M$ (in solar masses) approaching with velocity at infinity of 3 km/s the Solar System 
(model B). The units of $b$ and $\theta$ are those adopted in this paper.}
\label{tab:conc}
\end{table}

\section{Acknowledgments}

D. Punzo thanks the Leiden Observatory (University
of Leiden) for a period of hospitality.  This work was made possible also thanks to a financial 
contributions from the Dept. of Physics (Sapienza, University of Rome), from the Netherlands 
Research Council NWO (grants \#643.200.503, \#639.073.803 and \#614.061.608), from the Netherlands
Research School for Astronomy (NOVA), and from the HPC-EUROPA2 project
(project number: 1249) with the support of the European Commission -
Capacities Area - Research Infrastructures.  Most of the computations
were carried out on the computers owned by the ASTRO research group (Dep. of Physics, Sapienza, 
Univ. of Roma) and on the Little Green Machine at Leiden University and on the Lisa cluster at 
SURFSara in Amsterdam and by.  We finally thank M. Spera for his help in porting the code to 
various architectures.

\bibliographystyle{mn2e}
\bibliography{KBO.bib}

\begin{thebibliography}{}

\bibitem[\protect\citeauthoryear{{Aarseth}}{{Aarseth}}{2003}]{ars}
{Aarseth} S.~J.,  2003, {Gravitational N-Body Simulations}

\bibitem[\protect\citeauthoryear{{B{\'e}dorf} \& {Portegies
  Zwart}}{{B{\'e}dorf} \& {Portegies Zwart}}{2012}]{Jeroen}
{B{\'e}dorf} J.,  {Portegies Zwart} S.,  2012, European Physical Journal
  Special Topics, 210, 201

\bibitem[\protect\citeauthoryear{{Berczik}, {Nitadori}, {Zhong}, {Spurzem},
  {Hamada}, {Wang}, {Berentzen}, {Veles} \& {Ge}}{{Berczik}
  et~al.}{2011}]{Berczik1}
{Berczik} P.,  {Nitadori} K.,  {Zhong} S.,  {Spurzem} R.,  {Hamada} T.,  {Wang}
  X.,  {Berentzen} I.,  {Veles} A.,    {Ge} W.,  2011, in International
  conference on High Performance Computing, Kyiv, Ukraine, October 8-10, 2011.,
  p. 8-18 {High performance massively parallel direct N-body simulations on
  large GPU clusters.}.
pp 8--18

\bibitem[\protect\citeauthoryear{{Berczik}, {Spurzem} \& {Wang}}{{Berczik}
  et~al.}{2013}]{Berczik2}
{Berczik} P.,  {Spurzem} R.,    {Wang} L.,  2013, in Third International
  Conference ''High Performance Computing'', HPC-UA 2013, p. 52-59 {Up to 700k
  GPU cores, Kepler, and the Exascale future for simulations of star clusters
  around black holes.}.
pp 52--59

\bibitem[\protect\citeauthoryear{{Bernstein}, {Trilling}, {Allen}, {Brown},
  {Holman} \& {Malhotra}}{{Bernstein} et~al.}{2004}]{Ber}
{Bernstein} G.~M.,  {Trilling} D.~E.,  {Allen} R.~L.,  {Brown} M.~E.,  {Holman}
  M.,    {Malhotra} R.,  2004, Astronomical Journal, 128, 1364

\bibitem[\protect\citeauthoryear{{Binney} \& {Tremaine}}{{Binney} \&
  {Tremaine}}{1987}]{binney}
{Binney} J.,  {Tremaine} S.,  1987, {Galactic dynamics}

\bibitem[\protect\citeauthoryear{{Brown}}{{Brown}}{2001}]{Brown}
{Brown} M.~E.,  2001, Astronomical Journal, 121, 2804

\bibitem[\protect\citeauthoryear{{Brucker}, {Grundy}, {Stansberry}, {Spencer},
  {Sheppard}, {Chiang} \& {Buie}}{{Brucker} et~al.}{2009}]{Bru}
{Brucker} M.~J.,  {Grundy} W.~M.,  {Stansberry} J.~A.,  {Spencer} J.~R.,
  {Sheppard} S.~S.,  {Chiang} E.~I.,    {Buie} M.~W.,  2009, Icarus, 201, 284

\bibitem[\protect\citeauthoryear{{Capuzzo-Dolcetta}, {Mastrobuono-Battisti} \&
  {Maschietti}}{{Capuzzo-Dolcetta} et~al.}{2011}]{NBSymple}
{Capuzzo-Dolcetta} R.,  {Mastrobuono-Battisti} A.,    {Maschietti} D.,  2011,
  New Astronomy, 16, 284

\bibitem[\protect\citeauthoryear{{Capuzzo-Dolcetta} \&
  {Spera}}{{Capuzzo-Dolcetta} \& {Spera}}{2013}]{GPUcomp}
{Capuzzo-Dolcetta} R.,  {Spera} M.,  2013, Computer Physics Communications,
  184, 2528

\bibitem[\protect\citeauthoryear{{Capuzzo-Dolcetta}, {Spera} \&
  {Punzo}}{{Capuzzo-Dolcetta} et~al.}{2013}]{HiGPUs}
{Capuzzo-Dolcetta} R.,  {Spera} M.,    {Punzo} D.,  2013, Journal of
  Computational Physics, 236, 580

\bibitem[\protect\citeauthoryear{{Doressoundiram}, {Boehnhardt}, {Tegler} \&
  {Trujillo}}{{Doressoundiram} et~al.}{2008}]{Dore}
{Doressoundiram} A.,  {Boehnhardt} H.,  {Tegler} S.~C.,    {Trujillo} C.,
  2008, {Color Properties and Trends of the Transneptunian Objects}.
pp 91--104

\bibitem[\protect\citeauthoryear{{Fraser} \& {Kavelaars}}{{Fraser} \&
  {Kavelaars}}{2009}]{Fraser}
{Fraser} W.~C.,  {Kavelaars} J.~J.,  2009, Astronomical Journal, 137, 72

\bibitem[\protect\citeauthoryear{{Gomes}}{{Gomes}}{2003}]{Gomes}
{Gomes} R.~S.,  2003, Icarus, 161, 404

\bibitem[\protect\citeauthoryear{{Holman}}{{Holman}}{1995}]{Holman}
{Holman} M.~J.,  1995, in {Kinoshita} H.,  {Nakai} H.,  eds, 27th Symposium on
  Celestial Mechanics, {The Distribution of Mass in the Kuiper Belt}.
p.~116

\bibitem[\protect\citeauthoryear{{Hotelling}}{{Hotelling}}{1931}]{Hotelling}
{Hotelling} H.,  1931, Annals of Mathematical Statistics, 2, 360

\bibitem[\protect\citeauthoryear{{Ida}, {Larwood} \& {Burkert}}{{Ida}
  et~al.}{2000}]{Ida1}
{Ida} S.,  {Larwood} J.,    {Burkert} A.,  2000, The Astrophysical Journal,
  528, 351

\bibitem[\protect\citeauthoryear{{Ito} \& {Tanikawa}}{{Ito} \&
  {Tanikawa}}{2002}]{Ito}
{Ito} T.,  {Tanikawa} K.,  2002, Monthly Notices of the Royal Astronomical
  Society, 336, 483

\bibitem[\protect\citeauthoryear{{Jewitt} \& {Luu}}{{Jewitt} \&
  {Luu}}{1993}]{Jewitt}
{Jewitt} D.,  {Luu} J.,  1993, Nature, 362, 730

\bibitem[\protect\citeauthoryear{{Kenyon} \& {Luu}}{{Kenyon} \&
  {Luu}}{1999}]{Kenyon}
{Kenyon} S.~J.,  {Luu} J.~X.,  1999, Astronomical Journal, 118, 1101

\bibitem[\protect\citeauthoryear{{Kobayashi} \& {Ida}}{{Kobayashi} \&
  {Ida}}{2001}]{Ida2}
{Kobayashi} H.,  {Ida} S.,  2001, Icarus, 153, 416

\bibitem[\protect\citeauthoryear{{Kobayashi}, {Ida} \& {Tanaka}}{{Kobayashi}
  et~al.}{2005}]{Ida3}
{Kobayashi} H.,  {Ida} S.,    {Tanaka} H.,  2005, Icarus, 177, 246

\bibitem[\protect\citeauthoryear{{Levison}, {Morbidelli}, {Van Laerhoven},
  {Gomes} \& {Tsiganis}}{{Levison} et~al.}{2008}]{Levi}
{Levison} H.~F.,  {Morbidelli} A.,  {Van Laerhoven} C.,  {Gomes} R.,
  {Tsiganis} K.,  2008, Icarus, 196, 258

\bibitem[\protect\citeauthoryear{{Levison} \& {Stern}}{{Levison} \&
  {Stern}}{2001}]{Levi1}
{Levison} H.~F.,  {Stern} S.~A.,  2001, Astronomical Journal, 121, 1730

\bibitem[\protect\citeauthoryear{{Luu} \& {Jewitt}}{{Luu} \&
  {Jewitt}}{2002}]{Luu}
{Luu} J.~X.,  {Jewitt} D.~C.,  2002, Astronomy and Astrophysics, 40, 63

\bibitem[\protect\citeauthoryear{{Malhotra}}{{Malhotra}}{1993}]{Mal1}
{Malhotra} R.,  1993, Nature, 365, 819

\bibitem[\protect\citeauthoryear{{Malhotra}}{{Malhotra}}{1995}]{Mal2}
{Malhotra} R.,  1995, Astronomical Journal, 110, 420

\bibitem[\protect\citeauthoryear{{Malmberg}, {Davies} \& {Heggie}}{{Malmberg}
  et~al.}{2011}]{Malberg}
{Malmberg} D.,  {Davies} M.~B.,    {Heggie} D.~C.,  2011, Monthly Notices of
  the Royal Astronomical Society, 411, 859

\bibitem[\protect\citeauthoryear{{Marsden}}{{Marsden}}{1980}]{mpc}
{Marsden} B.~G.,  1980, Celestial Mechanics, 22, 63

\bibitem[\protect\citeauthoryear{{Melita}, {Larwood} \& {Williams}}{{Melita}
  et~al.}{2005}]{Melita}
{Melita} M.~D.,  {Larwood} J.~D.,    {Williams} I.~P.,  2005, Icarus, 173, 559

\bibitem[\protect\citeauthoryear{{Nitadori} \& {Makino}}{{Nitadori} \&
  {Makino}}{2008}]{nitadori}
{Nitadori} K.,  {Makino} J.,  2008, New Astronomy, 13, 498

\bibitem[\protect\citeauthoryear{{Noll}, {Grundy}, {Stephens}, {Levison} \&
  {Kern}}{{Noll} et~al.}{2008}]{Noll}
{Noll} K.~S.,  {Grundy} W.~M.,  {Stephens} D.~C.,  {Levison} H.~F.,    {Kern}
  S.~D.,  2008, Icarus, 194, 758

\bibitem[\protect\citeauthoryear{{Nyland}, {Harris } \& {Prins}}{{Nyland}
  et~al.}{2007}]{nyland}
{Nyland} L.,  {Harris } M.,    {Prins} J., , 2007, Fast N-Body Simulation with
  CUDA

\bibitem[\protect\citeauthoryear{{Pelupessy}, {J{\"a}nes} \& {Portegies
  Zwart}}{{Pelupessy} et~al.}{2012}]{HUYANO}
{Pelupessy} F.~I.,  {J{\"a}nes} J.,    {Portegies Zwart} S.,  2012, New
  Astronomy, 17, 711

\bibitem[\protect\citeauthoryear{{Pelupessy}, {van Elteren}, {de Vries},
  {McMillan}, {Drost} \& {Portegies Zwart}}{{Pelupessy} et~al.}{2013}]{amuse2}
{Pelupessy} F.~I.,  {van Elteren} A.,  {de Vries} N.,  {McMillan} S.~L.~W.,
  {Drost} N.,    {Portegies Zwart} S.~F.,  2013, Astronomy and Astrophysics,
  557, A84

\bibitem[\protect\citeauthoryear{{Portegies Zwart} \& {Boekholt}}{{Portegies
  Zwart} \& {Boekholt}}{2014}]{Tjarda}
{Portegies Zwart} S.,  {Boekholt} T.,  2014, ArXiv e-prints

\bibitem[\protect\citeauthoryear{{Portegies Zwart}, {McMillan}, {Harfst},
  {Groen}, {Fujii}, {Nuall{\'a}in}, {Glebbeek}, {Heggie}, {Lombardi}, {Hut},
  {Angelou}, {Banerjee}, {Belkus}, {Fragos}, {Fregeau}, {Gaburov} \&
  {Izzard}}{{Portegies Zwart} et~al.}{2009}]{amuse1}
{Portegies Zwart} S.,  {McMillan} S.,  {Harfst} S.,  {Groen} D.,  {Fujii} M.,
  {Nuall{\'a}in} B.~{\'O}.,  {Glebbeek} E.,  {Heggie} D.,  {Lombardi} J.,
  {Hut} P.,  {Angelou} V.,  {Banerjee} S.,  {Belkus} H.,  {Fragos} T.,
  {Fregeau} J.,  {Gaburov} E.,    {Izzard} R.,  2009, New Astronomy, 14, 369

\bibitem[\protect\citeauthoryear{{Portegies Zwart}, {McMillan}, {van Elteren},
  {Pelupessy} \& {de Vries}}{{Portegies Zwart} et~al.}{2013}]{simon2}
{Portegies Zwart} S.,  {McMillan} S.~L.~W.,  {van Elteren} E.,  {Pelupessy} I.,
     {de Vries} N.,  2013, Computer Physics Communications, 183, 456

\bibitem[\protect\citeauthoryear{{Portegies Zwart}}{{Portegies
  Zwart}}{2009}]{simon1}
{Portegies Zwart} S.~F.,  2009, The Astrophysical Journal, 696, L13

\bibitem[\protect\citeauthoryear{{Portegies Zwart}, {Belleman} \&
  {Geldof}}{{Portegies Zwart} et~al.}{2007}]{Simon}
{Portegies Zwart} S.~F.,  {Belleman} R.~G.,    {Geldof} P.~M.,  2007, New
  Astronomy, 12, 641

\bibitem[\protect\citeauthoryear{{Press}, {Teukolsky}, {Vetterling} \&
  {Flannery}}{{Press} et~al.}{2002}]{recipes}
{Press} W.~H.,  {Teukolsky} S.~A.,  {Vetterling} W.~T.,    {Flannery} B.~P.,
  2002, {Numerical recipes in C++ : the art of scientific computing}

\bibitem[\protect\citeauthoryear{{Volk} \& {Malhotra}}{{Volk} \&
  {Malhotra}}{2011}]{Volk}
{Volk} K.,  {Malhotra} R.,  2011, The Astrophysical Journal, 736, 11

\end{thebibliography}

\end{document}